\documentclass[apj]{emulateapj}

\newcommand{\HII}{H {\small II}  }
\newcommand{\kms}{{\rm ~km~s}^{-1}}
\newcommand{\Msun} {M_\sun}
\newcommand{\accrate} {M_\sun~{\rm yr}^{-1}}
\newcommand{\mjyb}{{\rm ~mJy~beam}^{-1}}
\newcommand{\cmcub}{cm$^{-3}$}

\newcommand{\jybkms}{{\rm Jy~beam}^{-1}{\rm ~km~s}^{-1}}
\newcommand{\tothe}{^}

\shorttitle{HIERARCHICAL ACCRETION IN G20.08-0.14 N}
\shortauthors{GALV\'AN-MADRID ET AL.}

\begin{document}


\title{Formation of an O-Star Cluster by Hierarchical Accretion in G20.08-0.14 N}


\author{Roberto Galv\'an-Madrid\altaffilmark{1,2,3}, Eric Keto\altaffilmark{1}, 
Qizhou Zhang\altaffilmark{1}, \\ Stan Kurtz\altaffilmark{2}, Luis F. 
Rodr{\'\i}guez\altaffilmark{2}, and Paul T. P. Ho \altaffilmark{1,3}}

\email{rgalvan@cfa.harvard.edu}


\altaffiltext{1}{Harvard-Smithsonian Center for Astrophysics, 60 Garden Street, 
Cambridge MA 02138, USA}
\altaffiltext{2}{Centro de Radioastronom{\'\i}a y Astrof{\'\i}sica, Universidad 
Nacional Aut\'onoma de M\'exico, Morelia 58090, Mexico}
\altaffiltext{3}{Academia Sinica Institute of Astronomy and Astrophysics, P.O. 
Box 23-141, Taipei 106, Taiwan}


\slugcomment{The Astrophysical Journal, 706, 1036, 2009 December 1}

\begin{abstract}
Spectral line and continuum observations of the ionized 
and molecular gas in G20.08-0.14 N explore the dynamics
of accretion over a range of spatial scales in this massive star-forming region.  
Very Large Array observations 
of NH$_3$ at $4"$ angular resolution show a large-scale (0.5 pc) molecular
accretion flow around and into a star cluster with three
small, bright \HII regions. 
Higher resolution ($0.4''$) observations with the 
Submillimeter Array in hot core
molecules (CH$_3$CN, OCS, and SO$_2$) and the 
VLA in NH$_3$, 
show that the two brightest and smallest \HII regions 
are themselves surrounded by smaller scale (0.05 pc)
accretion flows.  
The axes of rotation of the large and small scale flows 
are aligned, and the timescale for the contraction of the cloud 
is short enough, 0.1 Myr, for the large-scale accretion flow to
deliver significant mass to the smaller scales within the star formation
timescale. 
The flow structure appears to be continuous and hierarchical from larger to smaller scales.

Millimeter radio recombination line (RRL) observations at $0.4\arcsec$ angular resolution 
indicate rotation and outflow of the ionized gas within the brightest \HII region (A). 
The broad
recombination lines and a continuum spectral energy distribution (SED) 
that rises continuously from cm to mm wavelengths, are both characteristic of the class of 
\HII regions known as ``broad recombination line objects''. The SED 
indicates a density gradient inside this \HII region, and the RRLs 
suggest supersonic flows. These observations are consistent with
photoevaporation of the inner part of the rotationally flattened molecular 
accretion flow. 

We also report the
serendipitous detection of a new NH$_3$ (3,3) maser.
\end{abstract}


\section{Introduction} \label{intro}

Massive star-forming regions (MSFRs) with O stars
are usually identified by a group of
hypercompact (HC) \HII or ultracompact (UC) \HII regions 
found together, deeply
embedded in a dense molecular cloud \citep[][]{GL09,Church02,Hoar07}. 
That several \HII regions are typically found 
within each star-forming region
indicates that massive stars form together in small clusters. Furthermore,
the infrared luminosity and radio continuum brightness of the
individual \HII regions suggest that some of them 
may themselves contain 
more than one massive star. Thus, the spatial structure of massive 
star-forming regions
is clustered and hierarchical: the star-forming regions contain a
number of separate HC and UC \HII regions, each of 
which may in turn contain a few massive stars. 

Low angular resolution, single-dish,
molecular line surveys of MSFRs show evidence for large 
scale contraction of the embedding
molecular clouds \citep{WE03,KW08}. 
Higher angular resolution observations
of some of these regions
identify velocity gradients consistent with rotation and inflow.
In addition to the accretion flows seen
on the large-scale ($\sim 0.3-1$ pc) of the embedding molecular cloud 
(G10.6-0.4: Ho \& Haschick 1986, Keto et al. 1987a, Keto 1990; 
G29.96-0.02: Olmi et al. 2003), accretion flows are also
seen on smaller ($\leq 0.1$ pc) scales around individual HC and UC \HII regions 
(G10.6-0.4: Keto et al. 1988, Sollins et al. 2005a; W3(OH): Keto et al. 1987b, 
Keto et al. 1995; 
W51e2: Zhang \& Ho 1997, Young et al. 1998; G28.20-0.05: Sollins et al. 2005b; 
G24.78+0.08: Beltr\'an et al. 2004, 2005, 2006, Galv\'an-Madrid et al. 2008; 
G29.96-0.02: Beuther et al. 2007). 

It is unclear how the flows on different length scales are related. 
In the case of G10.6-0.4, the 
cluster-scale accretion flow can be traced down from the
largest cloud scale to the small scale of the brightest \HII region,
but it is not known whether this holds for other objects. For example,
in a survey of MSFRs, selected on the basis of IRAS colors and
specifically excluding those with \HII regions,
multiple bipolar molecular outflows (implying the presence of accretion flows) are seen 
in random orientations \citep{Beu02a,Beu02,Beu03}. The different orientations of these 
smaller-scale flows suggest separate, individual centers of collapse. 
This comparison raises the question whether a large-scale coherent flow is required for the
formation of the most massive stars, O stars ($M_\star > 20$ $\Msun$)
capable of producing bright \HII regions, whereas
B stars require only smaller scale flows.

It is also unclear what happens in an accretion flow when the inflowing 
molecular gas reaches the boundary of an embedded \HII region. 
Previous observations suggest that
the \HII regions in an MSFR that are surrounded by accretion flows,
may be best 
understood as deriving from the continuous ionization of the accretion flow 
\citep{Keto02, Keto03, Keto07}, rather than as a dynamically separate expanding bubble 
of ionized gas within the flow. 
Part of the ionized gas may continue 
to the central star or stars and part escapes off
the rotationally flattened accretion flow as a photoevaporative 
outflow perpendicular to the plane of rotation \citep{Hol94, YW96, Liz96, 
John98, Lugo04}. The outflow is accelerated to supersonic speeds by the density 
gradient maintained by the stellar gravity \citep{Keto07}. 
Because the extent of an ionized outflow is generally 
larger than the region of ionized inflow, in most cases the outflow should be detected 
more easily than the inflow. 
\HII regions classified as ``broad recombination line objects" (BRLO) 
\citep{JaffMP99,Sewi04} 
show steep density gradients and supersonic flows \citep{KZK08}, 
consistent with photoevaporation and acceleration. It is not known whether
all BRLO are associated with accretion. If the accretion surrounding an
O star cluster is continuous from the largest to the smallest scales, this
must be the case.

There are only a handful of radio recombination line (RRL) observations that spatially 
resolve the ionized flow within an HC \HII region. Velocity gradients consistent 
with outflow and rotation in the ionized gas have been previously reported for W3(OH) 
\citep{Keto95}, W51e2 \citep{KK08}, and G28.20-0.05 \citep{Sewi08}. 
Observations of the very massive and spatially 
large G10.6-0.4 \HII region made at the VLA in
the highest possible angular resolution are able
to map the inflowing ionized gas \citep{KW06}.

In order to study the accretion dynamics over a range of scales
in a MSFR, from the cluster scale
down to the scale of individual HC \HII regions and within the ionized gas, 
we set up a program of
radio frequency molecular line, recombination line, and continuum 
observations at two telescopes and with several different angular resolutions.
For this study we chose the massive star formation region G20.08-0.14 North
(hereafter G20.08N), 
identified by three UC and HC \HII regions detected in the cm continuum by 
\cite{WC89}. The total
luminosity of the region is $L \sim 6.6 \times 10^5$ $L_\odot$ for 
a distance of 12.3 kpc.\footnote{
Both near and far kinematic distances have been reported for G20.08N.  
The near value given by \cite{Dow80} ($d \approx 4.1$ kpc) 
is the most commonly quoted in the previous literature. 
In contrast, \cite{Fish03} and \cite{AnBan09} report  
that this region is at the far kinematic distance ($d \approx 12.3$ kpc). 
We will assume the far distance throughout the rest of the paper.  
For reference, a scale of $0.5\arcsec$ 
corresponds to $\approx 6000$ AU (0.03 pc). 
The total luminosity of the region was estimated to be   
$L \sim 7.3 \times 10^4$ $L_\odot$  
assuming the near kinematic distance \citep{WC89}. 
Correcting for the location at the far 
distance, the luminosity is $L \sim 6.6 \times 10^5$ $L_\odot$.}

Previous observations suggest accretion in the G20.08N cluster.
Molecular-line observations show dense gas embedding 
the \HII regions \citep{Tur79,Plume92}. 
Molecular masers, generally associated with
ongoing massive-star formation, have been detected in
a number of studies (OH: Ho et al. 1983; H$_2$O: Hofner \& Churchwell 1996; 
and CH$_3$OH: Walsh et al. 1998). 
\citet{KW07,KW08}  
observed large-scale inward motions consistent
with an overall contraction of the embedding molecular cloud. 
Those authors also observed SiO line profiles suggestive of 
massive molecular outflows, further evidence for accretion and star formation.  
The recombination line spectra show broad lines \citep{Gar85, Sewi04}, 
presumably due to large, organized motions in the ionized gas. 
However, the previous observations do not have the angular resolution and the range 
of spatial scales needed to confirm the presence of accretion flows and 
study them in detail.

In this paper we report on several observations of G20.08N and discuss our findings. 
We confirm active accretion within the cluster. Furthermore, we find that the parsec-scale 
accretion flow fragments into smaller flows around 
the individual HC \HII regions, and  that 
the gas probably flows from the largest scale down 
to the smallest scale. 
This continuous and hierarchical accretion may be necessary 
to supply enough mass to the small-scale flows to form O-type stars, 
in contrast to 
low- and intermediate-mass star-forming regions with stars 
no more massive than $\sim 20$ $\Msun$, where isolated accretion flows 
around individual protostars may be 
sufficient.

\section{Observations} \label{obs}

\subsection{SMA} \label{sma}

We observed  G20.08N on 2006 June 25 and July 6 with the
Submillimeter Array\footnote{The Submillimeter Array is a joint project between the Smithsonian 
Astrophysical Observatory and the Academia Sinica Institute of Astronomy and Astrophysics 
and is funded by the Smithsonian Institution and the Academia Sinica.} 
\citep{Ho04} in its very
extended (VEX) configuration.  
Two sidebands 
covered the frequency ranges of $220.3-222.3$ GHz and $230.3-232.3$ GHz 
with a spectral resolution of $\approx 0.5$ $\kms$.  
The H$30\alpha$ recombination line ($\nu_0=231.9009$ GHz) was positioned in the upper 
sideband\footnote{In ``chunk" 20 of the SMA correlator setup}. 
The observations sampled baseline lengths from $\approx 50$ to $\lesssim 400$ k$\lambda$, 
sensitive to a range of spatial scales from $\approx 0.5\arcsec$ to $\approx 4.1\arcsec$. 

The visibilities of each observation were separately calibrated using the
SMA's data calibration program, MIR.  
Table 1 lists relevant information on the calibrators. 
We used quasars for the absolute amplitude scale as well as the 
time-dependent phase corrections and  frequency-dependent bandpass corrections. 
Inspection of the quasar fluxes and comparison with their historical flux densities 
in the SMA database\footnote{http://sma1.sma.hawaii.edu/callist/callist.html} suggest that flux calibration
is accurate to better than $20 \%$. 
The calibrated data were exported to MIRIAD for further processing and imaging. 
 
There were enough line-free channels in the 2-GHz passband to subtract the continuum in the $(u,v)$ domain. 
The line-free continuum was self-calibrated in phase, and the gain solutions were applied to the 
spectral line data. We list all the identified lines in Table 2. Figure 1 shows the 
continuum-free spectra across the entire sidebands at the position of the 1.3-mm peak.

To improve the sensitivity, the data were smoothed to a spectral 
resolution of 2 $\kms$. The rms noise in our natural-weighted maps, 
made from the combined observations of both days, is $\sim 2$ $\mjyb$ for the single-sideband 
continuum and $\sim 30$ $\mjyb$ per channel (2 $\kms$ wide) for the line data. 

\subsection{VLA} \label{vla}

Spectral line observations of NH$_3$ $(J,K)=(2,2)$ and (3,3) at two different 
angular resolutions were obtained with the 
Very Large Array\footnote{
The National Radio Astronomy Observatory is operated by Associated Universities, Inc., 
under cooperative agreement with the National Science Foundation.}
in the D and BnA configurations (projects AS749, AS771, and AS785). 
Partial results from these observations were presented in \cite{SolPhD}. 
All the observations, except the VLA-D (3,3), were done with a 
bandwidth of 3.125 MHz ($\approx 39$ $\kms$) divided in 64 spectral channels, each 
$0.6$ $\kms$ wide. The VLA-D (3,3) observation was done with the same bandwidth divided 
in 128 spectral channels, each $0.3$ $\kms$ wide. 
The bandwidth covers the main hyperfine line and one line 
from the innermost satellite pair\footnote{The NH$_3$ molecule is symmetric top with 
inversion, see \cite{HT83} for details.}. 
The NH$_3$ data are presented at a spectral resolution of 0.6 $\kms$.  
The noise per channel in the final images was in the range of $1.0-1.5$ $\mjyb$.

In addition to the molecular line observations, we observed 
the ionized gas in the H66$\alpha$ recombination line in the VLA B 
configuration. The correlator was set up to cover a bandwidth of 12.5 MHz 
($\approx 166$ $\kms$) divided in 64 channels of 2.6 $\kms$ each. The rms noise 
per channel in the final image was $\approx 1$ $\mjyb$.

All three VLA data sets were calibrated using standard procedures in the AIPS software. 
Tables 1 and 2 summarize the relevant observational parameters.  
The continuum was constructed in the $(u,v)$ domain from line-free channels and was then 
self-calibrated. The gain solutions from self-calibration were applied to the line data.

\section{Results and Discussion} \label{res}

\subsection{The Continuum Emission} \label{cont}

\subsubsection{Morphology} \label{mor}

Figure  \ref{fig2} shows the 1.3-cm continuum (contours) obtained 
from the VLA-BnA  observations overlaid with the 1.3-mm continuum from the SMA-VEX 
data (color scale). At 1.3 cm we 
resolve the G20.08N system into the three components reported by \cite{WC89}. \HII region A 
is the brightest, westernmost peak. \HII region B is the slightly broader peak
$\approx 0.7\arcsec$ to the SE of A.
\HII region C is the more extended UC \HII further to the SE. Its brightest, 
eastern rim is detected at the $\sim 10$ $\mjyb$ level in our 1.3-mm observations.

The continuum of \HII region A is unresolved at 1.3 cm; at 1.3 mm it shows a core-halo 
morphology. The 1.3-mm core is unresolved (Gaussian fits yield a deconvolved size at 
half power FWHM $\lesssim 0.4\arcsec$). The low-intensity halo has a diameter of $\approx 1\arcsec$ 
(see Fig. \ref{fig2}). 
The H$30\alpha$ emission is confined to the unresolved core (\S~ \ref{ionized}) and the 
warm molecular gas (\S ~ \ref{smallscale}) coincides with the extended continuum halo. 
This indicates that the unresolved \HII region A is surrounded by a dust cocoon. 
\HII region B is barely resolved in the 1.3-cm VLA BnA map  
(deconvolved FWHM $\approx 0.6 \arcsec$). 
The peak position of \HII region A is identical at 1.3 cm and 1.3 mm: 
$\alpha\mathrm{(J2000)} = 18^{\mathrm h}$~$28^{\mathrm m}$~$10\rlap.{^{\mathrm s}}30$, 
$\delta\mathrm{(J2000)} = -11^{\circ}$~$28^\prime$~$47\rlap.{''}8$,  
within the positional uncertainty of the reference quasars (in the range 
$0.01\arcsec - 0.1\arcsec$).

\subsubsection{SED and the Nature of the Millimeter Emission} \label{sed}

Previous observations of
the recombination lines at 2 and 6 cm  \citep{Sewi04} put the \HII regions in G20.08N 
in the class known as ``broad recombination line objects'' (although those observations did 
not have sufficient angular resolution to separate \HII regions A, B, and C). The
large widths in cm-wavelength recombination lines 
are due to pressure broadening at high gas densities
($> 10^5$ cm$^{-3}$) as well as unresolved supersonic motions \citep{KZK08}. 
These \HII regions also have continuum spectral energy distributions (SEDs) that
increase with frequency through the mm wavelengths, evidence for a steep density gradient 
in the ionized gas \citep{Franco00, Keto03, Avalos06, Avalos09, KZK08}. 
Our new VLA and SMA observations, at 1.3 cm and 1.3 mm respectively, extend the SED to
millimeter wavelengths. We find that the flux density of  
\HII region A continues to rise from cm to mm wavelengths (Fig. \ref{fig2}), and we 
analyze this characteristic in detail below. 
The flux densities of \HII region B at 1.3 mm 
($\approx 93$ mJy) and 1.3 cm ($\approx 202$ mJy) imply a spectral index ($\alpha$, 
where $S_\nu \propto \nu\tothe\alpha$) 
of $\alpha\sim-0.3$, roughly consistent with the expected
index of $-0.1$ of optically thin gas. \HII region C is more extended and most 
of the 1.3-mm flux is resolved out.

At wavelengths shorter than 1 mm, thermal dust emission contributes 
significantly to the continuum. 
We estimate the relative contributions of dust and free-free
emission at 1.3 mm from our recombination line observations 
(described in \S ~ \ref{ionized}) and the theoretically 
expected line-to-continuum ratio.
In the optically thin limit (a good approximation at 
1.3 mm) the free-free line-to-continuum flux ratio $S_L/S_C$ is given by the 
ratio of the opacities 
$\kappa_L/\kappa_C$ \citep{GS02}: 

\begin{mathletters}
\begin{eqnarray}
\label{eq:kappal}
\biggl[\frac{\kappa_L}{\mathrm{cm}\tothe{-1}}\biggr] = \frac{\pi h\tothe3 e\tothe2}
{(2\pi m_e k)\tothe{3/2} m_e c} 
n_1\tothe2 f_{n_1,n_2} \phi_\nu \frac{n_e n_i}{T\tothe{3/2}}  \nonumber \\ 
\times \exp{\biggl(\frac{E_{n_1}}{kT}\biggr)}
\Bigl(1- e\tothe{-h\nu /kT}\Bigr),   ~ ~ ~ ~ ~ (1a) \nonumber \\
\biggl[\frac{\kappa_C}{\mathrm{cm}\tothe{-1}}\biggr] =  9.77 \times 10\tothe{-3} 
\frac{n_e n_i}{\nu\tothe2 T\tothe{3/2}}
\Biggl[17.72 + \ln{\frac{T\tothe{3/2}}{\nu}} \Biggr],  ~ ~ ~ ~ ~ (1b) \nonumber
\label{eq:kappac}
\end{eqnarray}
\end{mathletters}

\noindent
where all the units are in cgs, the physical constants have their usual meanings, $n_1=30$ 
for H $30\alpha$, 
$f_{n_1,n_2} \approx 0.1907n_1(1+1.5/n_1)$ for $\alpha$ lines, $n_e = n_i$ for hydrogen, 
$T \approx 8000-10$ 000 K, and $\phi_\nu$ is the normalized line profile. 
The main source of uncertainty in equations \ref{eq:kappal} and \ref{eq:kappac} is 
the temperature of the ionized gas. Assuming that the line profile is Gaussian 
and correcting for 8 $\%$ 
helium in the gas, the expected ratio at the line center is $S_{L,0}/S_C \approx 3.0$ 
for $T=8000$ K, or  $S_{L,0}/S_C \approx 2.3$ for $T = 10$ 000 K.  
The observed ratio is $S_{L,0}/S_C \approx 1.8$. Therefore, 
assuming that the RRLs are in LTE, the free-free 
contribution to the 1.3-mm flux of \HII region A 
is $\approx 60\%-80\%$ for the assumed temperature range.

Figure \ref{fig3} shows a model SED for \HII region A 
in which 70 $\%$ (355 mJy) of the 1.3-mm flux is produced by 
free-free from an \HII region with a density gradient
and 30 $\%$ (142 mJy) by warm dust. Assuming radiative equilibrium, we set the dust temperature 
$T_d$ to 230 K, the average temperature of the dense gas at the same scales (\S \ref{small-parms}).  
The modeling procedure is described in \citet{Keto03} and \citet{KZK08}.
Table 3 summarizes the model. 
The total gas mass inferred from the dust emission
is too large ($M \sim 35-95$ $\Msun$) for the \HII region alone, so most of the dust must be in 
the cocoon around the \HII region. The calculated mass range 
takes into account uncertainties in the dust emissivity, but not in the temperature. 

The density gradient derived for the ionized gas in \HII region A is  
$n_e\propto r\tothe{-\gamma}$, with $\gamma=1.3$. 
Equilibrium between recombination and ionization in this model \HII region
requires an ionizing flux equivalent to an O7.5 star \citep[using the computations of][]{Vacca96}, 
although this ionizing flux could
be made up of several stars of slightly later spectral type.
The model for the SED assumes spherical geometry and a static \HII region
with no inflow of neutral gas into the \HII region. In contrast, \HII region A
is embedded in a rotationally flattened accretion flow (see \S ~ \ref{smallscale}), 
so the determination of the stellar spectral type is only approximate.

\subsection{The Large-Scale Molecular Cloud} \label{vlad}

The large-scale molecular cloud is detected by the NH$_3$ VLA-D observations. 
From these data we find the presence of a parsec-scale accretion flow surrounding 
the cluster of \HII regions. In \S~ \ref{large-dyn} we first estimate the systemic velocity 
$V_{sys}$ of the cloud with respect to the local standard of rest (LSR), and then  
analyze the line velocities to determine rotation and infall following the procedure
used in measuring accretion flow velocities in previous papers 
\citep[e.g.,][]{HH86,KHH87,ZH97,Young98}. 
In \S~ \ref{large-parms} we 
derive the physical properties of the cloud: temperature, mass, ammonia abundance, and density.

\subsubsection{Dynamics} \label{large-dyn}

Figure \ref{fig4} shows the channel maps of the $(J,K)=(3,3)$ main hyperfine line. 
The most notable feature is a velocity gradient along the major axis of the 
cloud, consistent with rotation. The redshifted emission is toward the NE, while the blueshifted gas is 
toward the SW. Also, there is strong absorption against the free-free background of the \HII regions 
at the center of the cloud, which are unresolved at $\sim 4\arcsec$ resolution. The separation between 
the emission peaks on either side of the absorption is $\approx 10\arcsec$, or 0.6 pc.  
The symmetry of the channel maps suggests a systemic velocity of about $42$ $\kms$. 
A Gaussian fit to the emission spectrum integrated in a box covering the entire cloud, and clipping 
out the redshifted absorption, gives a centroid velocity of $42.3\pm0.4$ $\kms$. 
We adopt a systemic velocity for the parsec-scale 
cloud of $V_{sys}=42$ $\kms$, in agreement with that reported by \cite{Plume92} (41.9 $\kms$) 
based on observations of CS $J=7-6$.

Figure \ref{fig5} shows position-velocity (PV) diagrams 
across cuts at PA $=45\tothe \circ$ (SW-NE, major axis) and PA $=135\tothe \circ$ (NW-SE, 
minor axis). 
The rotation is seen in the SW-NE cut as a shift in the velocity 
of the emission contours from one side of the absorption to the other. 
The velocity offset with respect to $V_{sys}$ seen in the emission contours increases 
inward, suggesting that the gas rotates faster with decreasing radius.

Under the assumption that the velocity gradient seen in emission along 
the major axis is dominated 
by rotation, the redshifted absorption in the PV diagrams (Fig. \ref{fig5}) 
is also evidence for inward flow toward the 
central \HII regions, as there is an excess of redshifted absorption. 
This is more clearly seen in Fig. \ref{fig6}, which shows the spectra toward the absorption center. 
The NH$_3$ (2,2) main hyperfine absorption peak is redshifted by 2.3 kms$^{-1}$ with respect to the 
systemic velocity. An infall velocity $V_{inf} \approx 2$ $\kms$ is also seen in the NH$_3$ (3,3) 
line, 
although the spectrum is contaminated by an NH$_3$ (3,3) maser (see description in Fig. \ref{fig6}).
The maser is confirmed by our high angular resolution NH$_3$ observations (\S ~ \ref{33maser}).

If the rotation were seen edge-on, there would not be
a velocity gradient across the \HII region along the minor axis (NW-SE).
At an oblique viewing angle  a velocity gradient along the minor axis is 
created by the inflow.  
The velocity gradient along both the minor and major axes (Fig.~5)
implies that
the rotationally flattened flow is tipped with respect to the line of sight.

\subsubsection{Cloud Parameters} \label{large-parms}

The optical depth of the gas can be determined from the brightness ratios of 
the hyperfine lines.
From their optical depth ratio
the rotational temperature between the $(2,2)$ and $(3,3)$ transitions can be 
determined  \citep[see][for details of the procedure]{KHH87,Man92}.  
This temperature can be considered as a first-order  
approximation to the kinetic temperature ($T_k$) of the molecular gas.
\cite{Dan88} suggest that
an improved estimate of the kinetic temperature ($T_k$) is obtained by accounting
for the populations in the upper states of the K-ladders
rather than assuming that all states above the lowest
are negligibly populated. Including this correction,
we obtain an average kinetic temperature $T_k \sim 22$ K 
for the parsec-scale cloud.  
The high-resolution data discussed in \S ~ \ref{smallscale} show that higher
gas temperatures are found closer to the \HII regions. 

The mean column density of NH$_3$ is 
$N_{NH_3} \sim 2.3 \times 10^{16}$ cm$^{-2}$. We can determine
the abundance of ammonia $X(NH_3)$ by comparing the NH$_3$ and H$_2$ column 
densities.
The mass of the cloud  can be estimated from the observed velocity
dispersion ($\sigma_V \sim 3.5$ kms$^{-1}$), the radius of the cloud ($R \sim 100~000$ AU), 
and the virial theorem: $M_{vir} = (5/3G)R\sigma_V\tothe2 \sim 2300~\Msun$.
Similarly, from the observed velocity gradient, the rotation 
velocity is $V_{rot} \sim 4$ $\kms$ at a radius $R\sim 100~000$ AU. Equating the 
centripetal and gravitational forces, the gas mass inside $R$ is 
$M=(RV_{rot}\tothe2)/G \sim 1800$ $\Msun$.  
Therefore, we estimate the average column density of H$_2$ to be $N_{H_2} \sim 5 \times 10^{22}$ 
cm$^{-2}$ and $X(NH_3) \sim 5 \times 10^{-7}$, within a factor of three of the value estimated for 
G10.6--0.4 \citep{Keto90}. 
The mean H$_2$ density in the large-scale cloud is $n_{H_2} \sim 10^4$ \cmcub.

\subsection{Molecular Gas in The Inner 0.1 pc} \label{smallscale}

Our SMA Very Extended (VEX)  
and VLA-BnA observations provide a rich view of the 
molecular environment close to the \HII regions. 
The lines detected with the SMA that have peak intensities $>300 \mjyb$  
and are not blended with any other line are listed in the top part of Table 2.  
We follow a similar outline as for the large-scale cloud. 

\subsubsection{Dynamics} \label{small-dyn}

Figure \ref{fig7a} shows the channel maps of OCS $J=19-18$, SO$_2$ 
$J(K_a,K_b)=11(1,11)-10(0,10)$, CH$_3$CN $J(K)=12(3)-11(3)$, and CH$_3$CN $J(K)=12(4)-11(4)$. 
These maps show that
the line emission from these hot-core molecules
is considerably brighter around \HII region A.
All the molecules show a velocity gradient across this source,
from the southwest to northeast, in a similar  
orientation as the larger, cluster-scale flow.
We do not detect molecular emission around \HII region B.
At this high angular resolution we are not sensitive to 
brightness temperatures of less than $\sim 10$ K ($3\sigma$) for the mm lines.

Figure \ref{fig8} shows the velocity-integrated intensity (moment 0) and the intensity-weighted 
mean velocity (moment 1) maps of the four lines previously shown in Fig. \ref{fig7a}. The 
integrated emission of CH$_3$CN $12(4)-11(4)$ is brightest 
in front of the \HII region, while for the other molecules the brightness peak is 
slightly offset from the continuum.   
These differences
reflect the relative brightness of each molecule with respect to
the continuum emission of \HII region A. The velocity gradient across \HII region A is also seen. 
Figure \ref{fig9} shows the velocity-integrated (moment 0) and velocity dispersion 
(moment 2) maps for the same lines as Fig. \ref{fig8}. The line widths increase toward  
the continuum peak, indicating that unresolved motions increase closer to the 
\HII region.

Figure 10 shows the position-velocity (PV) diagrams for the lines of Figure 7 in
cuts at position angles PA $= 45^\circ$ and PA $= 135^\circ$ across the
continuum peak of HII region A. The cuts at $45^\circ$ show the 
velocity gradient also seen in the channel maps of OCS and 
both of the CH$_3$CN lines. The PV diagram of CH$_3$CN 12(4)-11(4) has a feature suggestive of 
a velocity gradient in the perpendicular direction PA $=135^\circ$, with an excess of redshifted 
emission toward the NW. 
Consistent with our interpretation of the NH$_3$ VLA-D data, this suggests 
inward motion in a rotationally-flattened flow that is seen not
quite edge-on. The same redshifted feature is also marginally detected in the lower excitation
CH$_3$CN transition as well as the OCS line. However, the infall signature in emission is only 
tentative, and a clearer indication of infall at small scales comes from the redshifted 
NH$_3$ absorption in the VLA-BnA data (see below). 

In general, observational experience suggests that 
CH$_3$CN, along with NH$_3$, is a 
reliable tracer of high-density molecular mass 
and accretion flows 
\citep{Cesa97,Zhang98,Zhang02,Patel05}. 
CH$_3$CN has recently been detected in the outflow
of the nearby low-mass star L1157 \citep{Arce08}, but at a very low brightness
($<0.03$ K). The distribution of the OCS molecule in our observations is very similar to 
the CH$_3$CN, but the SO$_2$ velocities do not show the same pattern, and are more difficult to 
interpret. 
The SO$_2$ may be more easily affected by the excitation conditions, 
and part of the observed emission could arise from the shocked boundaries 
of outflows. 
From Gaussian fits to the CH$_3$CN $K=2,3,$ and 4 emission 
lines at the position of the continuum peak, the systemic velocity at small scales is 
estimated to be $V_{sys} = 41.8 \pm 0.3~\kms$.  
Figure \ref{fig11} shows the CH$_3$CN spectra and their Gaussian fits.

At subarcsecond angular resolution, our NH$_3$ observations   
are sensitive to emission of brightness temperature above $\sim 200$ K. Therefore the 
thermal NH$_3$ is detected only in absorption against the bright continuum. 
As in \S ~ \ref{large-parms}, a comparison between the NH$_3$ absorption line velocity and
$V_{sys}$ shows an inward velocity of 
$\approx 2~\kms$ in front of \HII region A
(Figure \ref{fig12}). 
The NH$_3$ (2,2) absorption line in front of \HII region B is redshifted by 
$\sim 2.0$ $\kms$ with respect to $V_{sys}$, implying inward motion
and accretion toward \HII region B as well (Figure \ref{fig12}). 
The (3,3) absorption in front of \HII region B is mixed with NH$_3$ 
maser emission (\S ~ \ref{33maser}), and the determination of the inward velocity
is uncertain. More sensitive observations are needed to constrain the properties of 
the molecular gas around \HII region B. 

\subsubsection{Core Parameters} \label{small-parms}

We estimate the dynamical mass $M$ (gas plus stars) within the smaller 
accretion flow in the same way as with the large-scale flow (see \S ~ \ref{large-parms}). 
At a radius $R\sim 5000$ AU the rotation velocity is $V_{rot} \sim 3-4$ $\kms$. 
Therefore, $M\sim 50-90~\Msun$. This is consistent with the lower limit to the 
stellar mass, $M_\star \approx 35$ $\Msun$, required for ionization 
equilibrium (\S ~ \ref{sed}). The estimate is also consistent with the gas mass derived 
from the mm continuum once the free-free contribution has been properly 
subtracted, $M_{gas} \sim 35 - 95~\Msun$ (\S \ref{sed}). 
The mean H$_2$ density is $n_{H_2} \sim 10\tothe{6}$ \cmcub.

We derive the temperature in the dense gas surrounding \HII region 
A from the rotational energy diagram of the lines of the  
CH$_3$CN $J=12-11$ K-ladders \citep[see e.g.,][]{LorMun84, Zhang98}.
Figure \ref{fig13} shows this diagram for two cases: one 
considering all the $K=0,...,7$ lines, 
and the other including only the $K=4,...,7$ lines, which have lower optical depths  
than the low-number K lines 
(also, the $K=0,1$ lines are blended, separated by only 5.8 $\kms$). 
The rotational temperature obtained for the former case is $T_{rot} \sim 403$ K, while 
for the latter it is $T_{rot} \sim 230$ K. 
The difference between the two values given above appears 
to be caused by optical depth effects (the rotational diagram analysis assumes that the 
emission is optically thin). Although our sensitivity level does 
not permit us to detect the 
lines of the isotopologue CH$^{13}$CN 
and measure the optical depth of the 
CH$_3$CN emission, the upper limits are not restrictive ($\tau_{CH_3CN}<20$). High optical 
depths are also suggested by the flat slope of the K ladders (Fig. 1). 
Fitting the K ladders taking 
into account the opacities \citep{QZ09} yields a kinetic 
temperature close to the lower estimation, $T_k\sim225$ K, and optical depths 
are $>>1$  (Keping Qiu 2009, personal communication). 
The rotational temperatures are higher if we use only the brightest
pixels instead of averaging over all the emission. This suggests a temperature gradient 
toward the center of the \HII region.

We derive a CH$_3$CN column density of  $N_{CH_3CN} \sim 7.5 \times 10^{15}$ cm$^{-2}$ 
assuming $T_{k} \sim 230$ K.
Comparison of the CH$_3$CN column density with the dynamical
mass implies an abundance
$X$(CH$_3$CN) $\sim 5 \times 10^{-9} - 2 \times 10^{-8}$ 
for the range of masses quoted above. 
Abundance estimates in other MSFRs cover a range of values:
$\sim 10^{-10}$ inside the Orion hot core
and $\sim 10^{-11}$ outside \citep{LorMun84}; 
$10^{-8}$ 
inside the Orion hot core 
and $10^{-9}-10^{-10}$ in the Orion ridge
\citep{Wil94}; 
$\sim 3 \times 10^{-8}$ 
in Sgr B2(N) 
\citep{Num00};
$\sim 1-5 \times 
10^{-7}$ in W51e1/e2
\citep{Remij04}.

The optical depth of NH$_3$ in absorption 
toward \HII region A is
$\tau_{2,2} \sim 4.2$, $\tau_{3,3} \sim 0.9$.
The mean column density 
is $N_{NH_3}\sim 3.6\times 10^{16}$ 
cm$^{-2}$.  If the ammonia abundance at these small scales is in the range 
$X(NH_{3})=2\times 10\tothe{-6} - 1 \times 10\tothe{-7}$, then the molecular 
hydrogen column density is 
$N_{H_2} \sim 2 \times 10^{22}- 4 \times 10\tothe{23}$ cm$^{-2}$. 
This implies a molecular gas mass of $M_{gas} \sim 2-40$ $\Msun$.  
The kinetic temperature obtained from the NH$_3$ is  $T_{k} \sim 50$ K, 
considerably cooler than that obtained for CH$_3$CN and implying that most
of the NH$_3$ column density is further away from the \HII region
than the CH$_3$CN. 

The distribution of the warm molecular gas around \HII region A appears 
to be rotationally flattened (Figs. \ref{fig7a}, \ref{fig8}, \ref{fig9}), but   
the observed size is slightly larger than the radius at which the accretion flow is expected 
to become centrifugally supported (i.e., a ``disk''),   
$R_d = GM/V_{rot}\tothe2 \sim 2000 - 3000$ AU for a star of 
mass $M_\star=35~\Msun$ and $V_{rot}=3-4~\kms$. 
Actually, the disk scale matches the size of the \HII region (Table 3, \S \ref{ionized}).

The mass-inflow rate toward \HII region A 
can be estimated from the high-resolution NH$_3$ absorption. 
From an inflow velocity of $V_{inf} \approx 2~\kms$ and spherical geometry, 
the mass-inflow rate is $\dot{M} \sim 1 \times 10^{-3} - 2 \times 10\tothe{-2}~\Msun$ 
yr$^{-1}$ (for the $2-40~\Msun$ of molecular gas detected in ammonia absorption). 
This estimate may be an upper limit because the hot molecular core is
flattened rather than spherical.

\subsection{The Ionized Gas: Radio Recombination Lines} \label{ionized}

In \S ~ \ref{sed} we inferred a density gradient inside the HC \HII region A.  
In this section, we derive the internal dynamics of this \HII region based on multifrequency 
RRLs. The mm/sub-mm lines are especially important because they are much less 
affected by pressure broadening and preferentially trace denser gas. 
While subarcsecond resolution studies at wavelengths longer than 7 mm have been available 
for many years \citep{DePree97,DePree04,Rod08}, 
similar studies at shorter wavelengths have had limited angular resolution 
\citep[e.g.,][]{JaffMP99}. 
\cite{KZK08} presented the first results of high-frequency,
high-resolution ($\sim 1\arcsec$), multifrequency RRL 
observations in a sample of 5 MSFRs with similar characteristics to G20.08N A. 
They were able to separate the contributions of pressure broadening and large-scale motions to the 
line width, even when the \HII regions were unresolved. 
We follow their procedure to analyze our RRL data.

We observed the H30$\alpha$ ($\nu_0=231.90$ GHz) and H66$\alpha$  
($\nu_0=22.36$ GHz) lines at subarcsecond angular resolution (see Table 2). 
Because both the line-to-continuum ratio and the continuum intensity are lower at 22 GHz than 
at 231 GHz, the H66$\alpha$ line is much weaker than the H$30\alpha$ line. This 
is somewhat alleviated by the better sensitivity of the VLA, but the signal-to-noise ratio 
(S/N)in the high-frequency line is still better. 
Figure \ref{fig14} shows the moment 0 and moment 1 maps of the H$30\alpha$ line. 
Although the emission is unresolved at half power, there is a slight indication of a velocity 
gradient in the ionized gas that agrees (not perfectly) with the rotation seen in CH$_3$CN and OCS. 
Figure \ref{fig15} shows the H$30\alpha$ and H$66\alpha$ 
spectra toward \HII region A and their Gaussian fits. The H$66\alpha$ line 
shows evidence of a blueshifted wing, suggesting either inflow or outflow in addition
to rotation. 
Within the uncertainties of the fits, 
both the H66$\alpha$ and H30$\alpha$ lines have the same line width (Table 4).

Assuming that the dynamical broadening $\Delta v_D$ (caused by turbulence and ordered motions) 
and the thermal broadening $\Delta v_T$ are Gaussian, and that the pressure broadening 
$\Delta v_L$ is Lorentzian, the RRL has a Voigt profile with line width \citep{GS02}: 

\begin{eqnarray}
\Delta v_V(\nu) \approx 0.534\Delta v_L(\nu) + (\Delta v_D^2 + \Delta v_T^2 + 0.217\Delta v_L^2(\nu))^{1/2}, \nonumber 
\\ (2)
\nonumber 
\label{eq:Voigt}
\end{eqnarray}

\noindent
where all the widths are FWHM. 

For the H$30\alpha$ line at 231.9 GHz the pressure broadening is less than the
thermal broadening
at electron densities below $2\times 10^8$ cm$^{-3}$ \citep{KZK08}.  
Our SED modeling indicates lower densities over most of the HII region.
Therefore, the observed  line width can be attributed 
to thermal plus dynamical broadening.
The electron temperature $T_e$ in UC \HII regions is 
typically $T_e=8000-10$ 000 K, with a small gradient as a function of Galactocentric 
radius \citep[]{Aff96}. We adopt $T_e = 9000$ K 
($T_B< 100$ K for H30$\alpha$ because of the low optical depth and $\lesssim 1$ filling 
factor), which translates into a thermal FWHM of 
$\Delta v_T = 20.9~\kms$. Therefore, from Eq. \ref{eq:Voigt}, we obtain a  
dynamical width of $\Delta v_D = 24.8 ~\kms$. 
From the velocity gradient (Fig. \ref{fig14})  
it is seen that $\sim 6$ $\kms$ of $\Delta v_D$ can be in the form of rotation. 
The rest could be caused by inflowing or outflowing ionized gas, as suggested by 
the blueshifted ($\sim 2-3$ $\kms$) mean velocities of the RRLs, and 
by the blue wing in the H$66\alpha$ spectrum (Fig. \ref{fig15}).

Most of the ionized gas that we see should be outflow.
Inflow inside the \HII region is expected within the radius where the
escape velocity from the star exceeds the sound speed of the
ionized gas. This is approximately the Bondi-Parker transonic
radius  \citep{Keto07}, $R_b = GM_\star/2c_s^2 = 5.5$ AU $M_\star/\Msun$, or 
about 190 AU for \HII region A, 
assuming a sound speed $c_s=9$ kms$^{-1}$ and stellar mass $M_\star=35~\Msun$. 
\HII region A extends out to $\sim 2500$ AU, so most of
the gas is not gravitationally bound to the star and flows
outward. In this model, the outflow is continuously 
supplied by photoevaporation off the rotationally
flattened accretion flow (T. Peters et al. 2009, in preparation). The somewhat misaligned 
velocity gradient in the ionized gas (Fig.~14) derives from a combination of
the rotation and outflow blended together in the observing beam.

\subsection{Outflow Tracers} \label{co}

Low angular resolution (HPBW $=14\arcsec$), single-dish (JCMT)
observations of SiO (8-7) show evidence for 
large-scale molecular outflows in G20.08N \citep{KW08}.
Although the standard outflow tracers $^{12}$CO $J=2-1$ 
and $^{13}$CO $J=2-1$ are in our passband, we do not detect any CO in emission. 
The $(u,v)$ coverage of our SMA-VEX observations is incapable of imaging structures 
larger than $\sim 4 \arcsec$; therefore, the CO emission from the 
molecular cloud and molecular 
outflow must be on larger scales.
$^{12}$CO and 
$^{13}$CO are seen in our data in absorption at the position of \HII region A, 
at several different velocities in the 
range $V_{LSR} = 42 - 84~\kms$.
Some CO absorption features are at the same velocities as the HI absorption features of 
\cite{Fish03}
and are therefore due to foreground gas that is not related to G20.08N but rather to intervening
Galactic spiral arms.

\subsection{A New NH$_3$ (3,3) Maser} \label{33maser}

A handful of NH$_3$  masers have been reported in the literature, always associated with 
massive star formation \citep[e.g.,][]{Walsh07}. Many of the known NH$_3$  masers are from 
non-metastable ($J>K$) transitions. 
The first clear detection of a metastable ($J=K$) 
NH$_3$ (3,3) maser was reported by \cite{ManWoot94} toward DR 21(OH). 
Most of the detections point toward a shock 
excitation origin for the population inversion, inasmuch as the maser spots are invariably 
associated with outflow indicators such as bipolar CO and/or SiO structures, 
class I methanol masers, and/or water masers \citep{ManWoot94,KraemJack95,ZhangHo95}. 

We report the serendipitous detection of a new NH$_3$ (3,3) maser toward G20.08N. The maser 
spot is relatively weak, and is spatially centered at 
$\alpha\mathrm{(J2000)} = 18^{\mathrm h}$~$28^{\mathrm m}$~$10\rlap.{^{\mathrm s}}346$, 
$\delta\mathrm{(J2000)} = -11^{\circ}$~$28^\prime$~$47\rlap.{''}93$, 
close in projection to the center of \HII region B. 
The maser spot is spatially unresolved even in uniform weighting maps of 
the VLA-BnA data (HPBW = $0.48\arcsec \times 0.26 \arcsec$, PA = $-0.8^\circ$). 
If the deconvolved source size is
limited to half the beam size, then the peak brightness temperature of the spot is constrained to 
$T_B > 7 \times 10^3$ K. The high intensity, together with the absence of 
similar emission in our (2,2) maps at high angular resolution, confirm the maser nature of the 
(3,3) emission. The spectral feature is also very narrow (Fig. \ref{fig16}), typical
of maser emission,
although it shows evidence of line wings. 
From a Gaussian fit to the line profile, the velocity of the maser is
$V_{maser} \sim 44.7~\kms$  (accurate only to  20 \% $ -$ 40 \%).  
The FWHM is $\sim 0.7~\kms$, after deconvolving the channel width of $0.6~\kms$.  
Owing to its position, it is probable that the maser is excited by \HII region B. 
We do not have sufficient data, however, to assert that it is excited in a shock. 

\section{Hierarchical Accretion in G20.08N} 
\label{discussion}

\subsection{The Observations}

One of the central questions in star formation is whether
star formation is ``bimodal", i.e., whether high-mass and low-mass
stars form in a different way \citep{ShuAL87}. Analysis of recent 
observations suggests that accretion flows 
around protostars of all masses can
be explained within the standard model of star formation
as a combination of a thin disk inside a rotationally
flattened envelope (Kumar \& Grave 2007; Molinari et al. 2008; E. Keto \& Q. Zhang 2009, 
in preparation). 
Of course, once a protostar gains the mass and
temperature of an O star, the formation of an \HII
region within the accretion flow introduces new phenomena 
\citep{Keto02,KW06,Keto07}. 
Yet not all molecular clouds form massive stars. In a comprehensive
survey, \cite{Solo87} found that clouds 
that form low-mass stars are uniformly distributed throughout 
the Galactic disk, but those that form the clusters of the 
most massive stars, O stars capable of producing significant \HII regions, 
are associated with the galactic spiral arms. If the 
formation process is very similar (with the addition of significant ionization for $M_\star>20 M_\odot$), 
then the difference may be found in the conditions in the molecular clouds.
The observations of G20.08N reported here and of G10.6-0.4 
reported previously \citep{Keto90,KW06} suggest that 
one difference is that the molecular clouds surrounding 
young clusters of O stars are in a state of overall collapse
whereas in star-forming regions without \HII regions we see
only localized collapse. 

In G20.08N the \HII regions A, B, and C are  
surrounded by a common molecular cloud of 
radius $\sim 0.5$ pc and mass $M\sim2000~\Msun$, 
which is rotating with a velocity of $\sim 4$ $\kms$
and contracting with an inward velocity of $\sim 2$ $\kms$. 
This rotation and contraction constitute a large-scale
accretion flow.
The velocity of the inward flow is about equal to 
the rotational velocity implying that the gas is 
approximately in free-fall and not constrained by 
centrifugal force.
Within this larger flow are at least two smaller accretion 
flows around \HII regions A and B. The molecular core 
around \HII region A is bright enough to be detected in hot-core 
molecules in emission at subarcsecond 
angular resolution. The radius and mass of this core are 0.05 pc and 
$\sim 20~\Msun$. Accretion in the core is indicated by
rotation at a velocity of 3 or 4 $\kms$ and contraction 
of $\sim 2$ $\kms$.  We do not detect molecular emission 
around \HII region B, but the accretion inflow is inferred from 
NH$_3$ absorption that is redshifted by $\sim 2~\kms$ with
respect to the systemic velocity.

In contrast, star-forming regions that contain only low 
mass stars or even stars as massive as type B do not appear to have   
this global collapse of the entire parent cloud. The 
observations of these regions suggest only localized accretion 
flows within individual clumps.
For example, star-forming regions
such as IRAS 19410+2336 \citep{Srid02,Beu02,Beu03}, 
IRAS 05358+3543 \citep{Beu02a,Leu07}, and AFGL 5142 
\citep{Zhang02,Zhang07}, have bolometric luminosities of 
at most a few times $10\tothe4$ $L_\odot$, consistent with
type B protostars. 
There are a number of cores of size similar to those in
G20.08N, but there is no reported evidence of a larger parsec-scale, 
accretion flow.

In an analysis of a recent numerical simulation, 
\cite{VS09} also find that
the formation of massive stars or clusters
is associated with large-scale collapse 
involving thousands of $\Msun$ and accretion
rates of $10^{-3}~\accrate$. In contrast,
low- and intermediate-mass stars or clusters in their simulation
are associated with isolated accretion flows that are a factor of
10 smaller in size, mass, and accretion rate.

If global collapse of the host molecular cloud is necessary 
for the formation of O stars in clusters, then the association of O
stars with galactic spiral arms may imply that
compression  of giant molecular clouds
as they pass through galactic spiral arms may be the primary mechanism for
initiating global collapse
\citep{Roberts69,Shu72,Martinez09}. 
The low-mass star-forming regions found by \cite{Solo87} 
to be spread throughout the Galaxy may not need such
large-scale compression.

\subsection{Resupply of the Star-Forming Cores}

The orientations of the large cloud-scale accretion flow 
and the core-scale flow around \HII region A are similar and
the flows could be continuous (we do not know the orientation 
of the flow around \HII region B because we cannot detect 
the surrounding molecular emission).  The molecular core 
around \HII region A contains only a few tens of $\Msun$, 
similar to the mass of a single O star. 
If the core is to form one or more O stars at less
than 100 $\%$ efficiency, its mass must be resupplied by the 
larger scale accretion flow. Resupply is also suggested by the 
short dynamical or crossing timescale, given by the ratio of the size to 
the infall velocity, $\sim 10^4$ yrs.  
If the core is to last more than this, it must be 
resupplied by the larger scale flow, which has a dynamical timescale of $\sim 10^5$ yrs. 
If the core is not resupplied then the growing protostar may
simply run out of gas before reaching the mass of an O star.

A similar process of resupply is suggested in recent theoretical work. 
Analytic arguments show that as an unstable cloud fragments, there should be a 
continuous cascade of mass from larger to smaller fragments 
as well as a cascade of kinetic energy \citep{Field08,NewWass90}. 
In recent numerical 
simulations of high mass star formation, \cite{Wang09} 
find that most of the mass is supplied from outside a 
0.1 pc core around the protostar. \cite{VS09} and T. Peters et al. (2009, in preparation) 
also find in their simulations that as a massive core is consumed 
from the inside by an accreting protostar, the core continues to accrete mass from the outside.

\subsection{Accretion Rate}

The accretion rate within the small-scale flow around \HII region A
is $\dot{M} \sim 1 \times 10^{-3}$ to $2 \times 10\tothe{-2}~\Msun$.
Similar accretion rates are reported for flows around other \HII regions 
such as G10.6-0.4, $8 \times 10^{-4}~\Msun$ yr$^{-1}$ \citep{Keto90}, and
W51e2, $3 \times 10^{-3}$  to  $1 \times 10^{-2}~\Msun$ yr$^{-1}$ \citep{Young98}. 
In contrast, the accretion rates estimated for the cores 
in these MSFRs without bright \HII regions are generally lower
by one or two orders of magnitude, $\sim 10^{-4}~\accrate$ \citep{Beu02a,Beu02}.

A high accretion rate is necessary to form an O star.
Because accreting massive protostars begin core hydrogen 
burning well before reaching the mass of an O star,
they evolve essentially as main-sequence stars of
equivalent mass. 
Numerical simulations of stellar evolution that include
accretion predict that unless the accretion rate is
high enough, a growing massive protostar will
evolve off the main-sequence and explode before it
reaches the mass of an O star \citep{KW06}. 
For a protostar to gain the mass of an O star, the rate at which accretion 
supplies fresh hydrogen to the growing protostar must be
greater than the rate at which the star burns the hydrogen.
At the upper end of the mass spectrum, $M_\star>40~\Msun$, 
this rate is $> 10^{-3}~\accrate$. 
The accretion flows in G20.08N are capable of supplying
gas at the rate necessary to form massive O stars.
In MSFRs without O stars, the accretion rates may be
too low for the stars to achieve the 
mass of an O star within their hydrogen-burning lifetimes.

\subsection{Transfer of Angular Momentum}

If the flow in G20.08N is continuous, then the 
observations show that the flow spins up as it contracts. 
Ignoring projection effects, the 
magnitude of the specific angular momentum on the large scale is
$L/m \sim 0.6~\kms$ pc at $r \approx 0.2$ pc, 
from the VLA-D NH$_3$  data, 
while from the VEX SMA data it is $L/m \sim 0.1~\kms$ pc 
at $r \approx 0.02$ pc around \HII region A.  
From these estimates $\sim 85$ \% of the specific 
angular momentum in the large-scale flow is lost.  
In previous observations of G10.6-0.4 (Keto 1990), we  found
that 97\% of the angular momentum in that flow
is lost between 1.5 and 0.02 pc.
Evidently, angular momentum is efficiently transferred
outward and does not prevent collapse of the cloud.

Magnetic fields may be  
important in this process \citep[e.g.,][]{Girart09}. Although there are no
observations of the magnetic field in G20.08N, the
field direction has been mapped in another
MSFR with HC \HII regions, W51e1/e2. Here, observations show that the
magnetic field is uniform on
the larger scale, 0.5 pc \citep{Lai01}, while
on the smaller scale, 0.03 pc, of the accretion flow onto
W51e2, the field is pinched into an hourglass shape
with the accretion flow at the waist \citep{Tang09}. 
Thus ordered field lines
extend from the \HII region-scale accretion flow
to the large-scale molecular cloud, and if the field
has enough strength, angular momentum could be
transferred outward by the field. However, 
dust polarization observations do not give a direct estimate of the 
field strength, and it is also
possible that the field is essentially passive
and just carried along by the flow. The observation
that the clouds in both W51e1/e2 and G20.08N
are close to free-fall collapse
implies that the magnetic field is not strong
enough to support
the clouds. In other words, these clouds are 
magnetically supercritical.

Numerical simulations of star formation that do
not include magnetic forces show that angular momentum 
can be transferred by hydrodynamics alone.
\cite{Abel02} find that at any radius, there is both low and high
angular momentum gas, and that pressure
forces or shock waves can redistribute the
angular momentum between fluid elements.
Lower angular momentum gas 
sinks inward and displaces higher
angular momentum gas outward, resulting in a
net outward flow of angular momentum.

Whether the specific angular momentum is transferred by
hydrodynamics or magnetic forces, observations and simulations show that 
angular momentum is not conserved as a function of radius, and does
not prevent the gas from flowing continuously from large to small scales in a rotating flow.

\section{Conclusions} 
\label{conclu}

We report radio and mm observations of the molecular and ionized gas toward the O-star cluster 
G20.08N, made with an angular resolution from $\sim 0.1$ pc to $\sim 0.01$ pc.  Our main findings 
can be summarized as follows: 

\begin{enumerate}

\item
We find a large-scale ($\sim 0.5$ pc) accretion flow around and into a star cluster with several O-type stars, 
identified by one UC and two HC \HII regions. This flow is rotating and infalling towards its center. 
The two HC \HII regions are surrounded by smaller accretion flows ($\sim 0.05$ pc), each of them with the 
signature of infall too. The brightest (toward \HII region A) is detected in mm emission lines, 
and rotates in concordance with the large-scale flow. 

\item
The similar orientations of the flows at small and large scales, as well as their dynamical timescales 
($\sim 10^4$ yrs and $\sim 10^5$ yrs respectively), and masses ($\sim 10~\Msun$ 
and $\sim 10^3~\Msun$ respectively), suggest that, if O stars are forming in G20.08N (as it is observed), then 
the smaller scales ought to be resupplied from the larger scales. 
The same result has been found in recent numerical simulations of massive star formation in clusters.  

\item
The brightest HC \HII region (A) has a rising SED from cm to mm wavelengths and broad hydrogen recombination lines. 
Both characteristics suggest density gradients and supersonic flows inside the \HII region. A tentative 
velocity gradient is detected in the recombination line emission of this source, suggesting rotation and 
outflow in the ionized gas at the innermost scales. \HII region A can be interpreted as the inner part 
of the surrounding molecular accretion flow, with the observed ionization being produced by photoevaporation.

\end{enumerate}

\acknowledgements

We thank the anonymous referee for a detailed report. 
We also thank Keping Qiu for his comments on the temperature determination. 
R. G.-M. acknowledges support from an SMA predoctoral fellowship.

\clearpage

\begin{center} 
\begin{deluxetable}{ccccccc}  \label{tab1}
\tabletypesize{\scriptsize}
\tablecaption{Observational Parameters}
\tablewidth{0pt}
\tablehead{
\colhead{Epoch} & \colhead{Array} & \multicolumn{2}{c}{Phase Center\tablenotemark{a}} & 
\colhead{Bandpass} & \colhead{Phase} & \colhead{Flux} \\\cline{3-4} 
 &  &  $\alpha$(J2000) & $\delta$(J2000) & Calibrator & Calibrator & Calibrator 
}
\startdata
2003 Apr 28 & VLA-D & 18 28 10.384 & $-11$ 28 48.65 & 3C454.3 & $1833-210$ & $0137+331$ \\
2003 May 13 & VLA-D & 18 28 10.384 & $-11$ 28 48.65 & 3C454.3 & $1851+005$ & $1331+305$ \\
2003 Oct 09 & VLA-BnA & 18 28 10.384 & $-11$ 28 48.65 & 3C454.3 & $1851+005$ & $1331+305$ \\
2003 Oct 10 & VLA-BnA & 18 28 10.384 & $-11$ 28 48.65 & 3C454.3 & $1851+005$ & $1331+305$ \\
2006 Jun 25 & SMA-VEX & 18 28 10.38 &  $-11$ 28 48.60 & 3C273 & $1830+063$ & $1830+063$ \\
2006 Jun 25 & SMA-VEX & 18 28 10.76 &  $-11$ 29 27.60 & 3C273 & $1830+063$ & $1830+063$ \\
2006 Jul 06 & SMA-VEX & 18 28 10.38 &  $-11$ 28 48.60 & 3C454.3 & $1751+096$ & $1751+096$ \\
2006 Jul 06 & SMA-VEX & 18 28 10.76 &  $-11$ 29 27.60 & 3C454.3 & $1751+096$ & $1751+096$ \\
2007 Oct 26 & VLA-B & 18 28 10.400 & $-11$ 28 49.00 & 3C454.3 & $1743-038$ & $1331+305$ \\
2007 Oct 27 & VLA-B & 18 28 10.400 & $-11$ 28 49.00 & 3C454.3 & $1743-038$ & $1331+305$ \\
\\
\hline
\enddata
\tablenotetext{a}{Units of right ascension are hours, minutes, and seconds. Units of declination are 
degrees, arcminutes, and arcseconds.}
\end{deluxetable}
\end{center}

\begin{center} 
\begin{deluxetable}{cccccc}  \label{tab2}
\tabletypesize{\scriptsize}
\tablecaption{Lines\tablenotemark{a}}
\tablewidth{0pt}
\tablehead{
\colhead{Species} & \colhead{Transition} & $\nu_0$  & Array & HPBW  \\
   &   & (GHz)  &   & (arcsec $\times$ arcsec; deg) 
}   
\startdata
H & 66$\alpha$ & 22.364178  & VLA-B & $0.47 \times 0.34$; 3 \\
NH$_3$ & $(2,2)$ & 23.722633  & VLA-D & $4.68 \times 2.93$; 236 \\
NH$_3$ & $(3,3)$ & 23.870129  & VLA-D & $4.50 \times 3.32$; 8 \\
NH$_3$ & $(2,2)$ & 23.722633  & VLA-BnA & $0.37 \times 0.28$; $343$ \\
NH$_3$ & $(3,3)$ & 23.870129  & VLA-BnA & $0.71 \times 0.34$; $0$ \\
$^{13}$CO & 2-1 & 220.398681   & SMA-VEX & $0.55 \times 0.41$; $32$  \\
CH$_3$CN & 12(4)-11(4) & 220.679297  & SMA-VEX & $0.55 \times 0.41$; $32$  \\
CH$_3$CN & 12(3)-11(3) & 220.709024  & SMA-VEX & $0.55 \times 0.41$; $32$  \\
CH$_3$CN & 12(2)-11(2) & 220.730266  & SMA-VEX & $0.55 \times 0.41$; $32$  \\
SO$_2$ & 11(1,11)-10(0,10) & 221.965200  & SMA-VEX & $0.54 \times 0.41$; $32$  \\
CO & 2-1 & 230.538000 &  SMA-VEX & $0.53 \times 0.39$; $37$  \\
OCS & 19-18 & 231.060991  & SMA-VEX & $0.53 \times 0.39$; $37$  \\
H & 30$\alpha$ & 231.9009   & SMA-VEX & $0.53 \times 0.39$; $37$  \\
\\
\hline
\\
CH$_3$CN & 12(7)-11(7) & 220.539340 & SMA-VEX & $0.55 \times 0.41$; $32$  \\
HNCO & 10(1,9)-9(1,8) & 220.584762 & SMA-VEX & $0.54 \times 0.41$; $32$  \\
CH$_3$CN & 12(6)-11(6) & 220.594438 & SMA-VEX & $0.55 \times 0.41$; $32$  \\
CH$_3$CN & 12(5)-11(5) & 220.641096 & SMA-VEX & $0.55 \times 0.41$; $32$  \\
CH$_3$CN & 12(1)-11(1) & 220.743015 & SMA-VEX & $0.55 \times 0.41$; $32$  \\
CH$_3$CN & 12(0)-11(0) & 220.747265 & SMA-VEX & $0.55 \times 0.41$; $32$  \\
CH$_2$CHCN & 24(0,24)-23(0,23) & 221.76598 & SMA-VEX & $0.55 \times 0.41$; $32$  \\
$^{13}$CS & 5-4 & 231.220768 & SMA-VEX & $0.52 \times 0.39$; $37$  \\
\\
\hline
\enddata
\tablenotetext{a}{Detected lines at S/N $>6$. The {\it top} part of the table lists the lines with 
S/N $>10$ (except the H$66\alpha$ line) and clearly isolated in frequency. 
The {\it bottom} part of the table lists the lines 
detected at S/N $<10$ or blended. Some spectral features (see Fig. \ref{fig1}) that were not properly 
identified due to low S/N ($\approx 5$) and blending are not listed.}
\end{deluxetable}
\end{center}


\clearpage

\begin{center}
\begin{deluxetable}{ccc} \label{tab3}
\tabletypesize{\scriptsize}
\tablecaption{Model for the \HII Region G20.08N A}
\tablewidth{0pt}
\tablehead{
\colhead{Parameter\tablenotemark{a}} &  & Value  
}
\startdata
HII Radius (AU) &  & 2530 \\
Electron density\tablenotemark{b} ($10^5$ \cmcub) &  & 1.4 \\
Exponent\tablenotemark{c} &  & 1.3 \\
Gas Mass\tablenotemark{d} ($\Msun$) &  & 35-95 \\
\cline{1-3}
Spectral Type &  & O7.5 \\
Stellar Mass ($\Msun$) &  & 34 \\
HII Mass ($\Msun$) &  & 0.05 \\
\hline
\enddata
\tablenotetext{a}{The independent parameters are the first four rows.}
\tablenotetext{b}{Electron density at the HII radius.}
\tablenotetext{c}{Exponent $\gamma$ of the power-law density gradient in the ionized gas, 
where $n \propto r^{-\gamma}$.}
\tablenotetext{d}{Mass of molecular gas obtained from the dust emission. 
Assuming dust temperature $T_d=230$ K and a gas-to-dust ratio of 100. 
The range is caused by 
the dust emissivity coefficient $\beta$ used \citep[see ][]{KZK08}, from $\beta=1$ 
to $\beta=1.5$.}
\end{deluxetable}
\end{center}


\begin{center} 
\begin{deluxetable}{ccccc}  \label{tab4}
\tabletypesize{\scriptsize}
\tablecaption{Emission Line Parameters Toward \HII Region A\tablenotemark{a}}
\tablewidth{0pt}
\tablehead{
\colhead{Species} & \colhead{Transition} & \colhead{$V_{LSR}$}  & \colhead{FWHM} & \colhead{$T_{B,peak}$}  \\
   &   & ($\kms$) &  ($\kms$) & (K)  
}   
\startdata
H & 66$\alpha$ & $39.3\pm1.1$ & $34.1\pm2.5$ & $190\pm13$ \\
H & 30$\alpha$ & $40.3\pm0.4$ & $32.5\pm0.9$ & $89\pm2$ \\
CH$_3$CN & 12(4)-11(4) & $42.7\pm0.3$ & $5.8\pm0.7$ & $32\pm4$ \\
CH$_3$CN & 12(3)-11(3) & $41.4\pm0.2$ & $5.3\pm0.5$ & $41\pm4$ \\
CH$_3$CN & 12(2)-11(2) & $41.4\pm0.2$ & $4.9\pm0.5$ & $32\pm3$ \\
\hline
\enddata
\tablenotetext{a}{From Gaussian fits. $1\sigma$ statistical errors are quoted.}
\end{deluxetable}
\end{center}

\clearpage

\begin{figure}
\figurenum{1}
\begin{center}
\includegraphics[angle=0,scale=0.4]{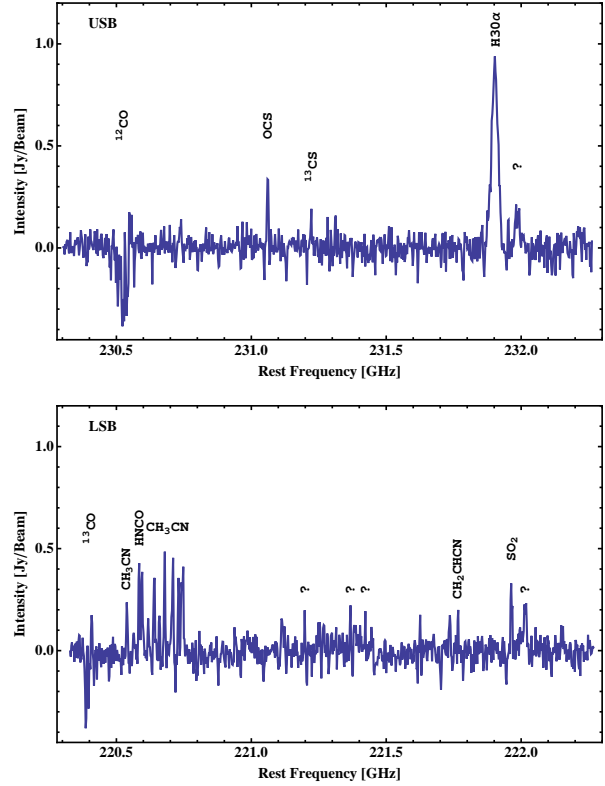}
\end{center}
\caption{
Wide-band, continuum-free spectra from the SMA-VEX data at the position of the 1.3-mm continuum peak 
(see Table 2 for details). The channel spacing in this plot is 3 $\kms$. 
The question mark ({\it ?}) in the upper sideband ({\it top} frame) 
might be a 
superposition of lines of CH$_3$OCH$_3$ and CH$_3$CH$_2$CN. The {\it ?s} at the center 
of the lower sideband ({\it bottom} frame) could be from vibrationally excited CH$_3$CN. 
The {\it ?} in the upper sideband close to the SO$_2$ might be from CH$_3$CCH. 
}
\label{fig1}
\end{figure}

\begin{figure}
\figurenum{2}
\begin{center}
\includegraphics[angle=0,scale=0.5]{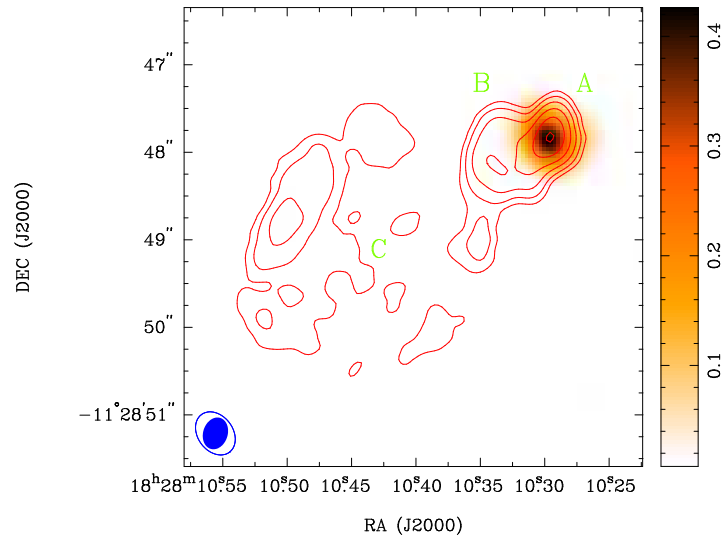}
\end{center}
\caption{
VLA-BnA 1.3-cm continuum ({\it red contours}) overlaid on the SMA-VEX 1.3-mm continuum 
({\it color scale}) 
toward  
the G20.08N complex. From west to east: \HII region A 
is the compact, strongest peak at both wavelengths. 
\HII region B is the less bright \HII region at $\lesssim 1 \arcsec$ to the SE of A. 
\HII region C is the more extended emission to the SE of B. 
The color scale goes linearly from 8 to 430 $\mjyb$ 
(the rms noise in the mm image is 2 $\mjyb$). 
Contours are placed at $-5,5,10,20,40,80,150$ $\times ~1$ $\mjyb$, the noise of the cm image.  
The SMA-VEX beam (empty ellipse) encircles the VLA-BnA beam (filled ellipse)
at the bottom-left of the image. 
}
\label{fig2}
\end{figure}

\clearpage

\begin{figure}
\figurenum{3}
\begin{center}
\includegraphics[angle=0,scale=0.6]{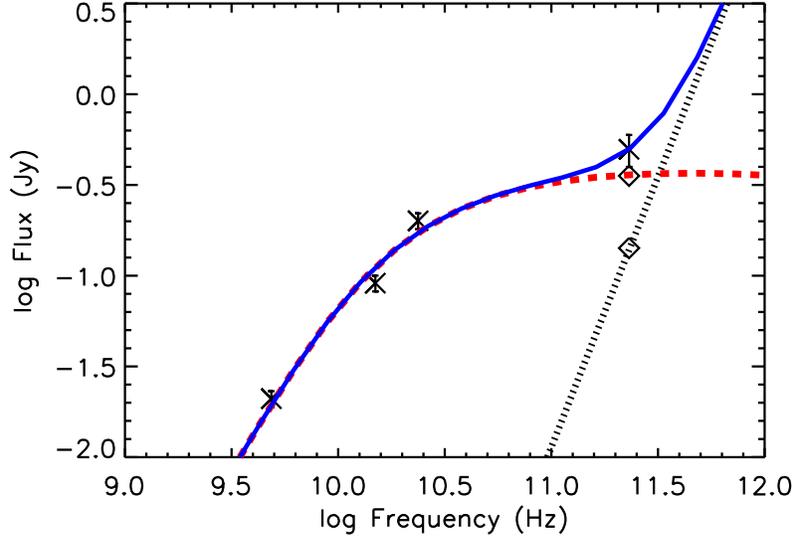}
\end{center}
\caption{
Radio-to-millimeter SED of G20.08N A.    
The 6-cm and 2-cm measurements were obtained from \cite{WC89}. 
The 1.3-cm and 1.3-mm points were obtained from Gaussian fits to our 
VLA-BnA and SMA-VEX data, respectively. {\it Crosses} are the 
data points. The error bars correspond to the $10\%$ and $20\%$ uncertainty 
expected in the VLA and SMA flux measurements, respectively. 
The {\it red dashed} line shows the flux of an \HII region with a density gradient. 
The {\it black dotted} line is the flux from the warm dust component. 
The {\it solid blue} line is the sum of the two components. 
The relative contributions of free-free (70 $\%$) and dust (30 $\%$) to the 
1.3-mm flux were estimated 
from the observed H$30\alpha$ line-to-continuum ratio, and are marked with 
{\it diamonds}.  
}
\label{fig3}
\end{figure}

\begin{figure}
\figurenum{4}
\begin{center}
\includegraphics[angle=0,scale=0.6]{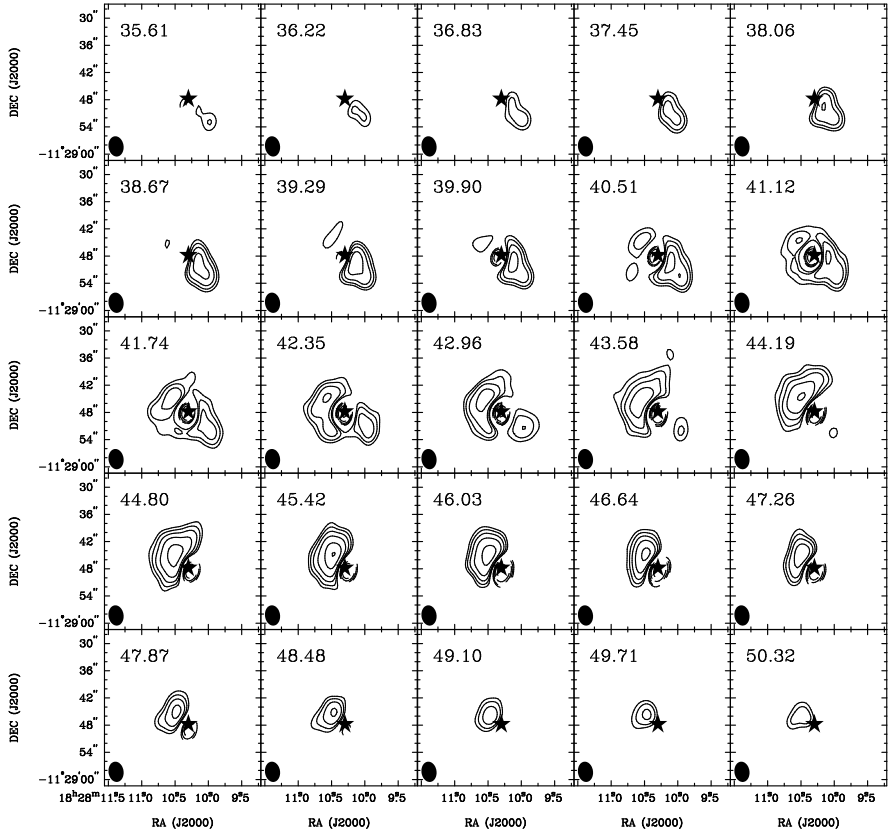}
\caption{
Channel maps of the VLA-D NH$_3$ (3,3) observations. Emission is in {\it solid} contours and 
absorption in {\it dashed} contours. The star covers the \HII regions shown in Fig. \ref{fig1}. 
Contour levels are at $-35,-25,-15,-10,-7,-5,5,7,10,15,25,35$ $\times ~2$ $\mjyb$. 
A clear velocity gradient in emission is seen from one side of the absorption 
to the other. The LSR systemic velocity of the molecular gas is $V_{sys} = 42.0$ 
$\kms$. The original maps at 0.3 $\kms$ spectral resolution were smoothed to 0.6 $\kms$ 
for clarity. 
}
\label{fig4}
\end{center}
\end{figure}

\clearpage

\begin{figure}
\figurenum{5}
\begin{center}
\includegraphics[angle=0,scale=0.35]{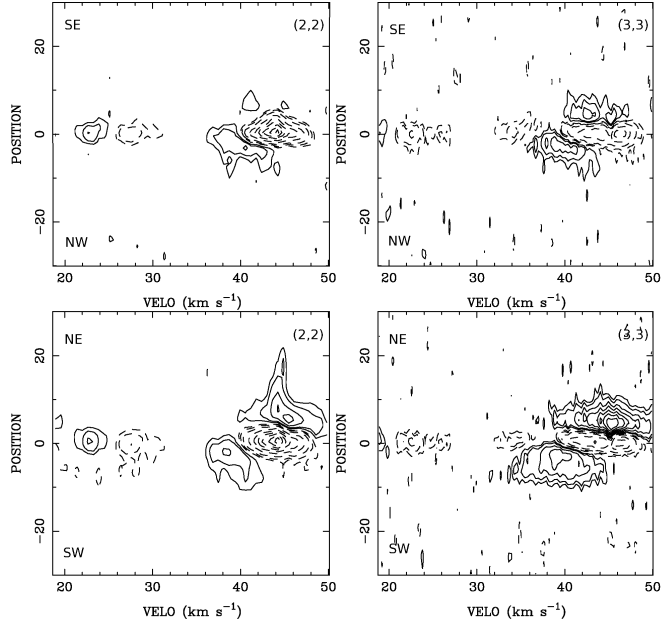}
\end{center}
\caption{Position-velocity diagrams of NH$_3$ (2,2) and (3,3) from the VLA-D data. 
The cuts were done at PA=45$^\circ$ ({\it bottom} row) and PA=135$^\circ$ ({\it top} row). 
{\it Dashed} contours are absorption, {\it solid} contours are emission. 
Contouring is at $-168,-144,-120,-96,-72,-48,-36,-24,-12,-4,4,8,12,16,24,32,40,48,56,64$ 
$\times ~1$ $\mjyb$. 
Only one inner satellite line is covered in the velocity range. 
In the SW-NE cuts (across the major axis of the cloud) the difference in the velocity of the 
emission with respect to $V_{sys}=42$ $\kms$ increases closer to the position center. 
This can be interpreted as spin up with decreasing distance from the center. 
However, the same trend is present in the NW-SE cuts (along the minor axis of the cloud), 
although only in the stronger, blueshifted side of the emission. This suggests 
that besides rotation, radial motions in the frame of the central stars are also present.}
\label{fig5}
\end{figure}

\begin{figure}
\figurenum{6}
\begin{center}
\includegraphics[angle=0,scale=0.6]{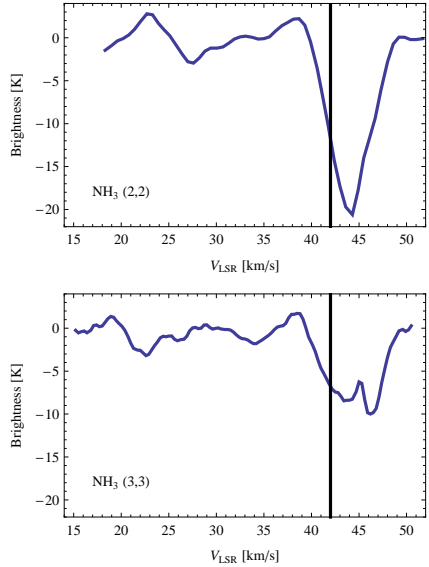}
\caption{
NH$_3$ spectra from the VLA-D observations toward the center of the absorption in G20.08N. The vertical line 
marks the systemic velocity ($V_{sys} = 42$ $\kms$). 
The small peak near the middle of the (3,3) absorption is due to maser emission (confirmed 
in the high-resolution, VLA-BnA data, see \S \ref{33maser}). 
The absorption peaks in the main 
(2,2) and (3,3) lines are redshifted with respect to $V_{sys}$, indicating the presence of  
inflow in the kinematics of the parsec-scale molecular cloud. 
The other weak absorption component seen in the  
(2,2) spectrum is one of the inner satellites. 
The absorption component at $\approx 22$ $\kms$ in the (3,3) spectrum is also an inner satellite. 
The weaker absorption at $30-37$ $\kms$ in the (3,3) spectrum may arise from 
an outflow. The blueshifted absorption in the main lines closest to $V_{sys}$ might arise from 
blending with the rotation seen in emission. 
}
\label{fig6}
\end{center}
\end{figure}

\clearpage

\begin{figure}
\figurenum{7}
\begin{center}
\includegraphics[angle=0,scale=0.5]{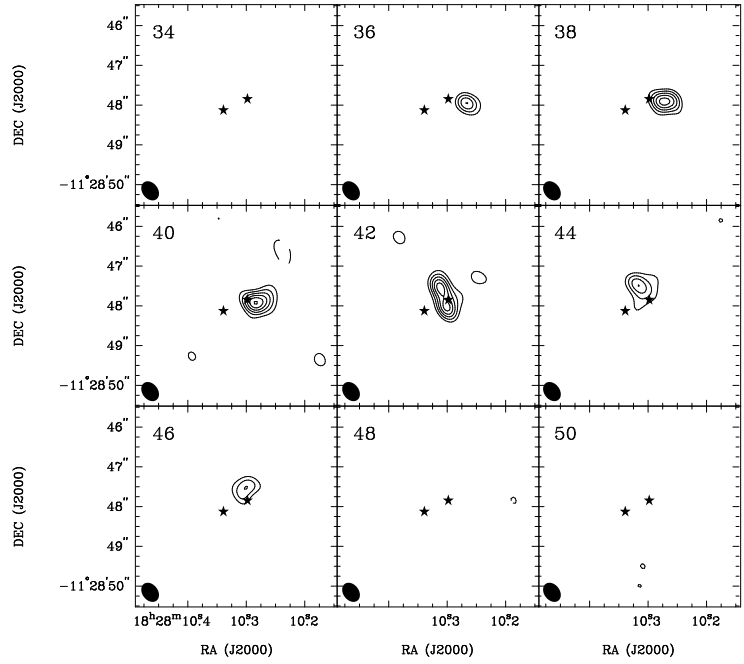}
\end{center}
\caption{Channel maps of OCS $J=19-18$ from the SMA-VEX observations. Contours are 
$-4,4,6,8,10,12,14$ $\times ~30~\mjyb$ (negative in $dashed$, and positive in $solid$). 
The peak intensity is $423~\mjyb$.  
The numbers in the upper left corner indicate the central LSR velocity of the channel. 
The two stars mark the positions of \HII regions A (west) and B (east). }
\label{fig7a}
\end{figure}

\begin{figure}
\figurenum{7 (continued)}
\begin{center}
\includegraphics[angle=0,scale=0.5]{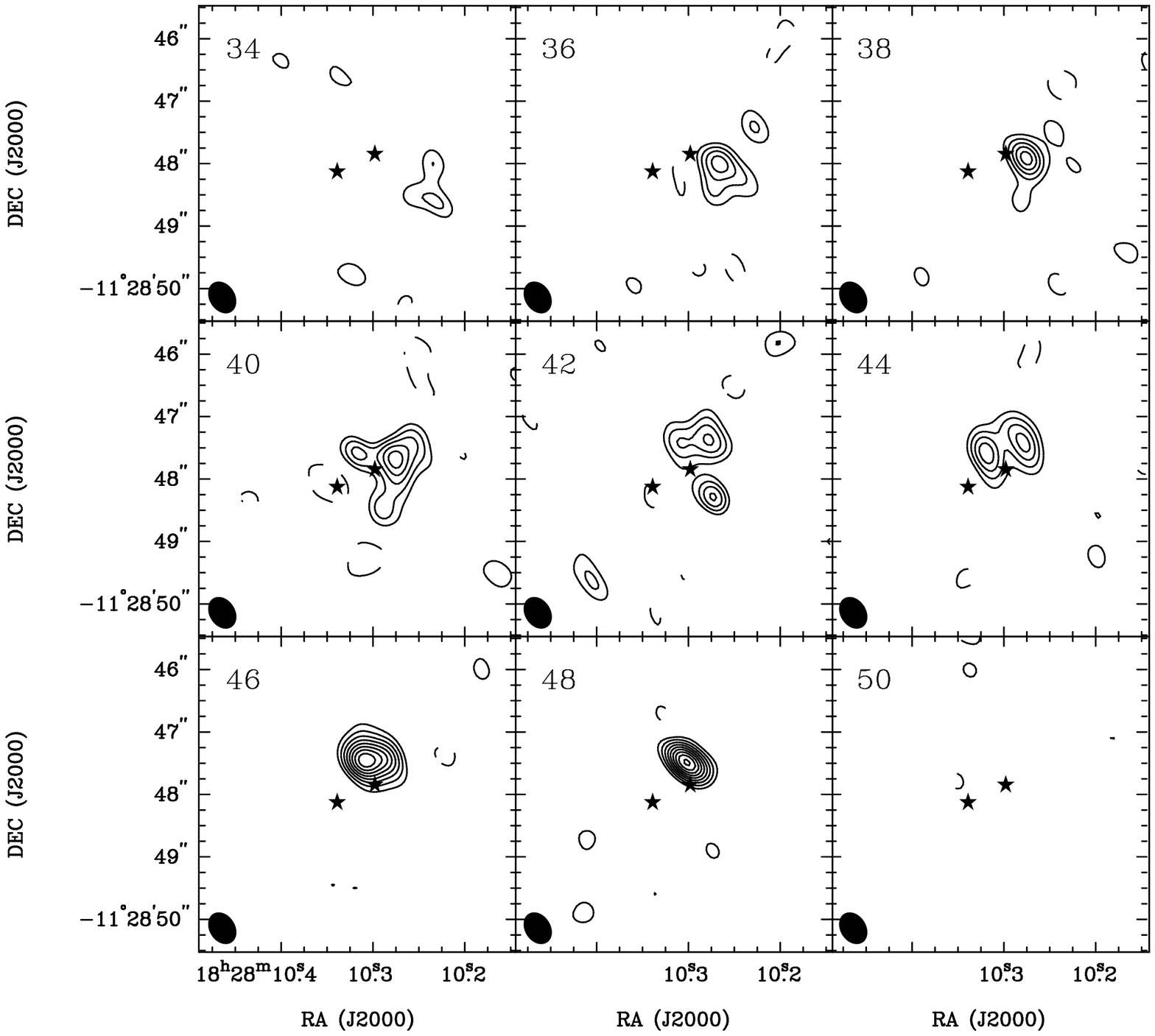}
\end{center}
\caption{Channel maps of SO$_2$ $J(K_a,K_b)=11(1,11)-10(0,10)$ from the SMA-VEX observations. 
Contours are $-4,4,6,8,10,12,14,16,18,20$ $\times ~30~\mjyb$ 
(negative in $dashed$, and positive in $solid$). The peak intensity is $612~\mjyb$.}
\label{fig7b}
\end{figure}

\clearpage

\begin{figure}
\figurenum{7 (continued)}
\begin{center}
\includegraphics[angle=0,scale=0.5]{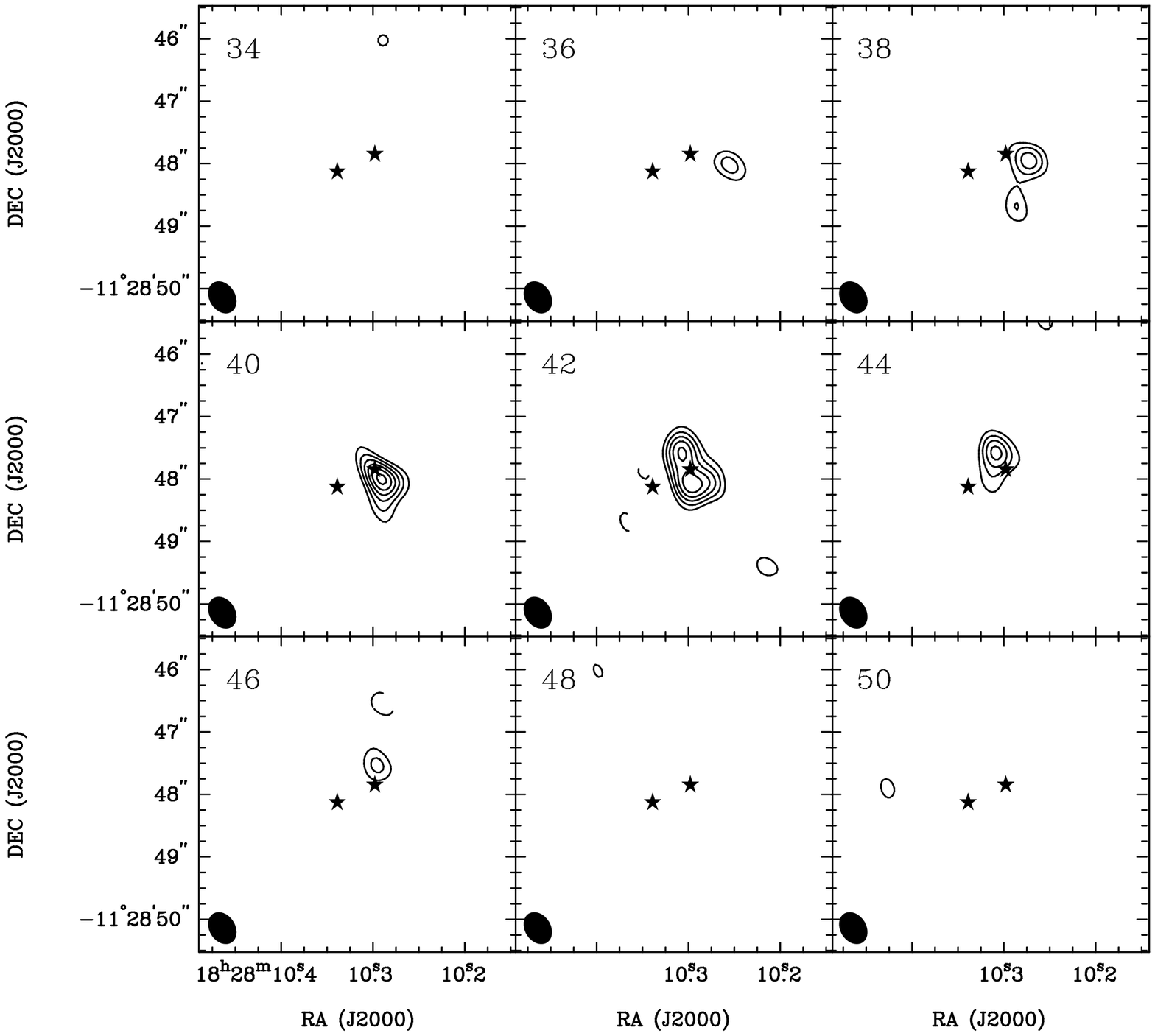}
\end{center}
\caption{Channel maps of CH$_3$CN $J(K)=12(3)-11(3)$ from the SMA-VEX observations. 
Contours are $-4,4,6,8,10,12,14$ $\times ~30~\mjyb$ (negative in $dashed$, and positive in $solid$). 
The peak intensity is $440~\mjyb$.}
\label{fig7c}
\end{figure}

\begin{figure}
\figurenum{7 (continued)}
\begin{center}
\includegraphics[angle=0,scale=0.5]{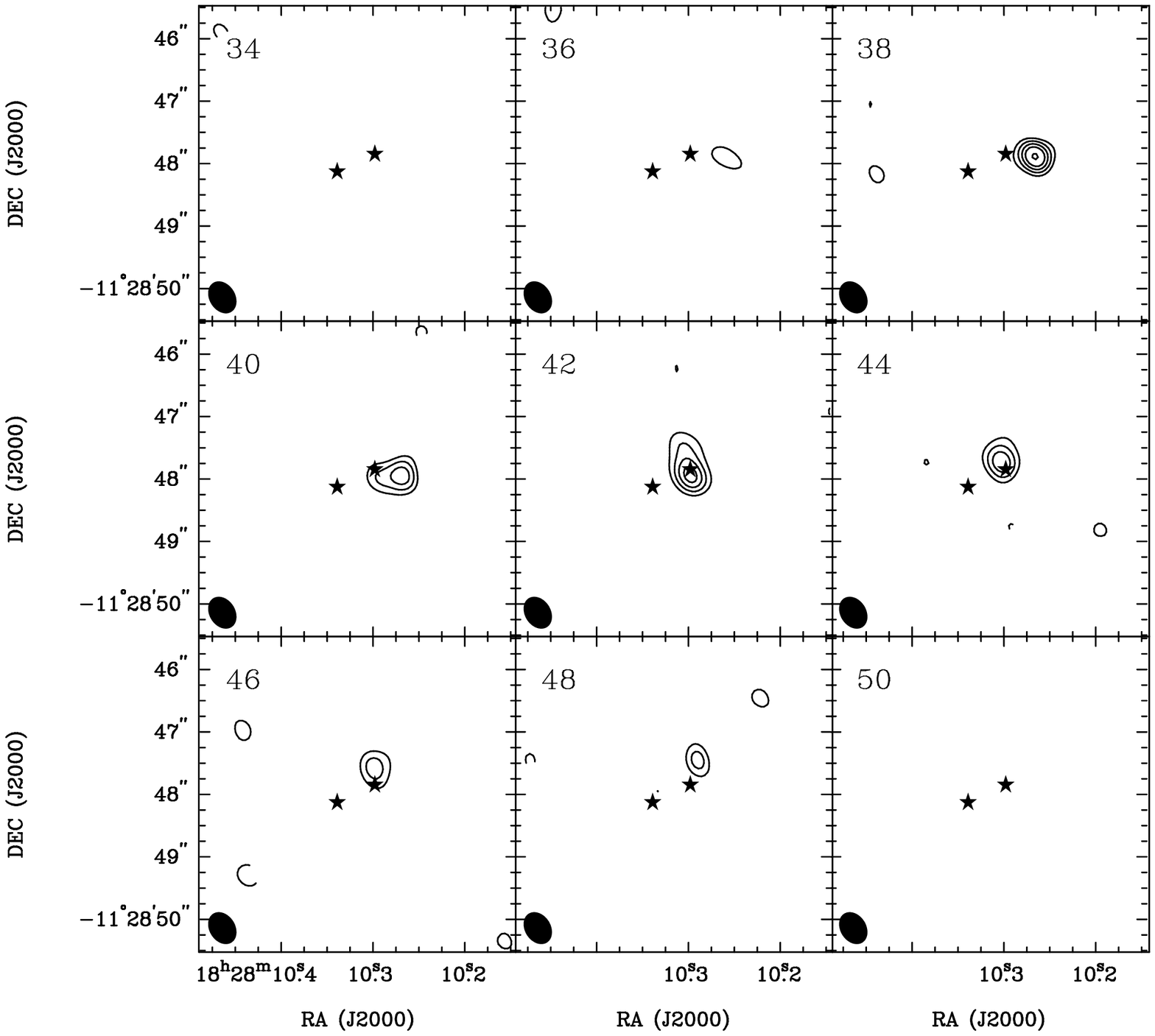}
\end{center}
\caption{Channel maps of CH$_3$CN $J(K)=12(4)-11(4)$ from the SMA-VEX observations. 
Contours are $-4,4,6,8,10,12,13$ $\times ~30~\mjyb$ (negative in $dashed$, and positive in $solid$). 
The peak intensity is $401~\mjyb$.}
\label{fig7d}
\end{figure}

\clearpage

\begin{figure}
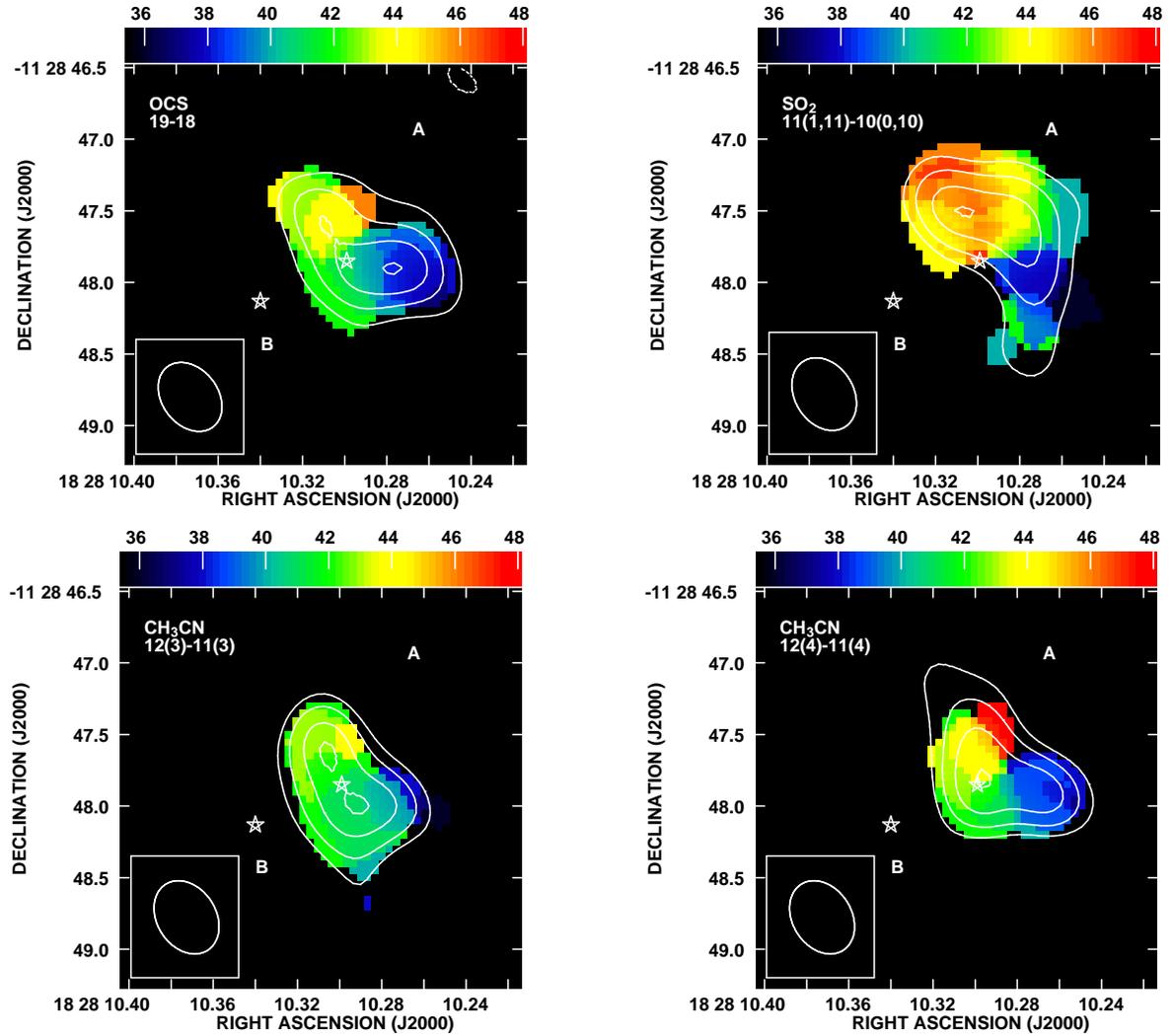

\figurenum{8}
\begin{center}
\plottwo{f8a.eps}{f8b.eps}
\plottwo{f8c.eps}{f8d.eps}
\caption{Velocity-integrated emission (moment 0, {\it contours}) and intensity-weighted mean velocity 
(moment 1, {\it color scale}) maps from hot-core molecules toward G20.08N. 
The {\it Top Left} panel shows the OCS $J=19-18$ (contours are $-5,5,9,13,17 \times 0.15~\jybkms$). 
The {\it Top Right} panel corresponds to the SO$_2$ $J(K_a,K_b)=11(1,11)-10(0,10)$ 
(contours are $-5,5,8,12,15 \times 0.25~\jybkms$). 
The {\it Bottom Left} panel shows the CH$_3$CN $J(K)=12(3)-11(3)$  
(contours are $-5,5,8,12,15 \times 0.15~\jybkms$). 
The {\it Bottom Right} panel plots the CH$_3$CN $J(K)=12(4)-11(4)$ 
(contours are $-5,5,9,14,19 \times 0.1~\jybkms$). 
Negative contours are in $dashed$ style, and positive contours in $solid$. 
Only the continuum \HII region A is associated with warm gas in 
our SMA-VEX data. The extent of the line emission is similar in all the tracers, 
although for the SO$_2$ it is more extended toward the northwest and more absorbed toward the continuum peak. 
The velocity gradient seen in the channel maps for each molecule is also seen here. 
The color scale is the same in all the frames.}
\end{center}
\label{fig8}
\end{figure}

\clearpage

\begin{figure}
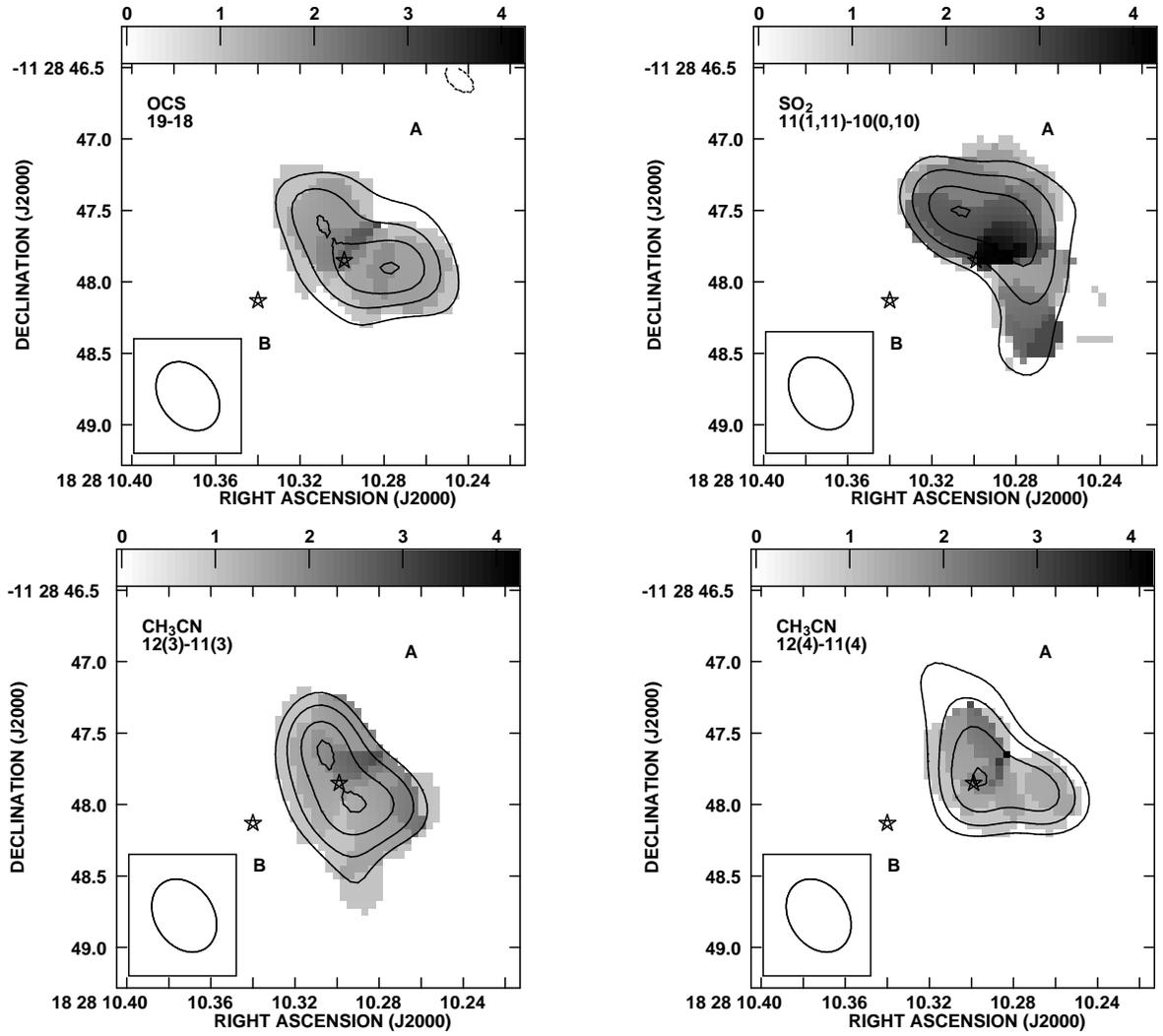

\figurenum{9}
\begin{center}
\plottwo{f9a.eps}{f9b.eps}
\plottwo{f9c.eps}{f9d.eps}
\caption{Velocity-integrated emission (moment 0, {\it contours}) and velocity dispersion with 
respect to moment-1 velocity (moment 2, {\it gray scale}) maps from hot-core molecules toward G20.08N. 
Contours are the same as in Fig. \ref{fig8}. The plotted velocity dispersion is 
$\sigma={\rm FWHM}/2(2\ln 2)^{1/2}$. $\sigma$ increases toward the 
center of \HII region A, possibly caused by 
unresolved motions toward the center.}
\end{center}
\label{fig9}
\end{figure}

\clearpage

\begin{figure}
\figurenum{10}
\epsscale{0.75}
\begin{center}
\plottwo{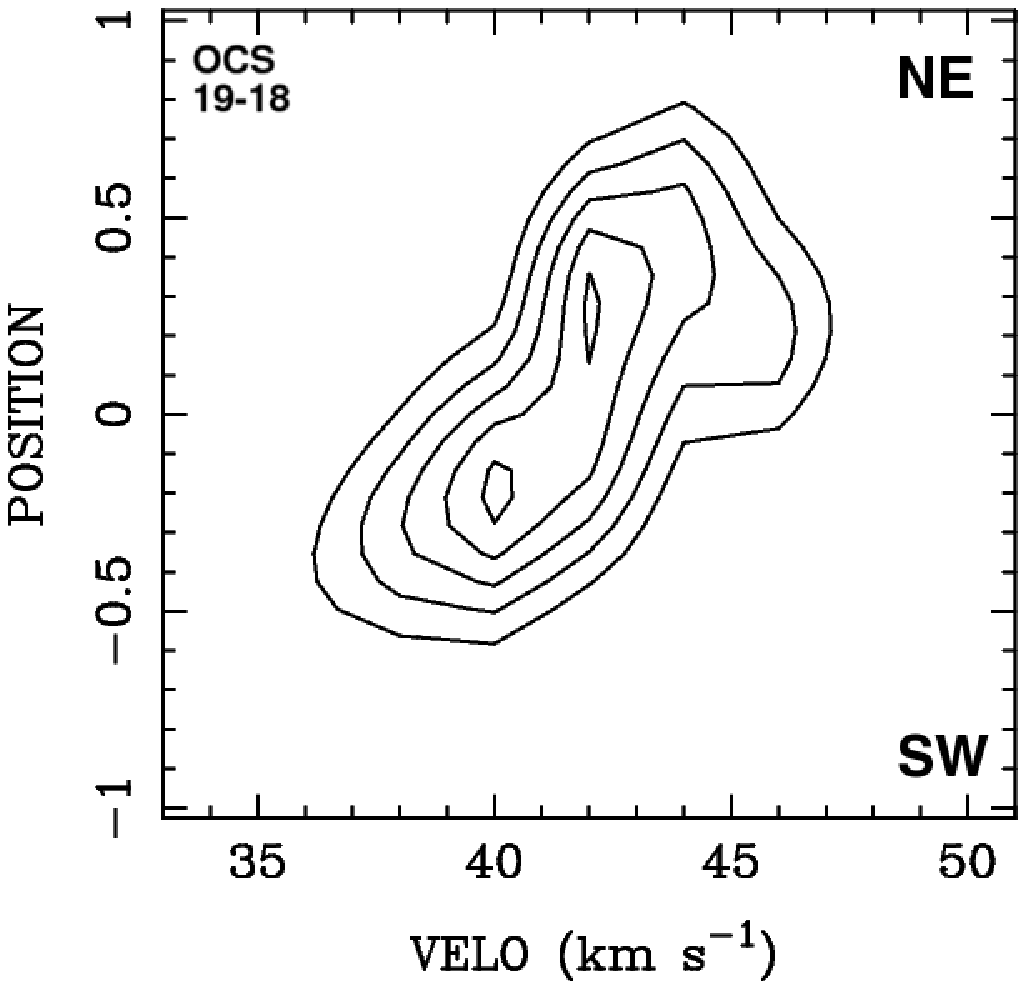}{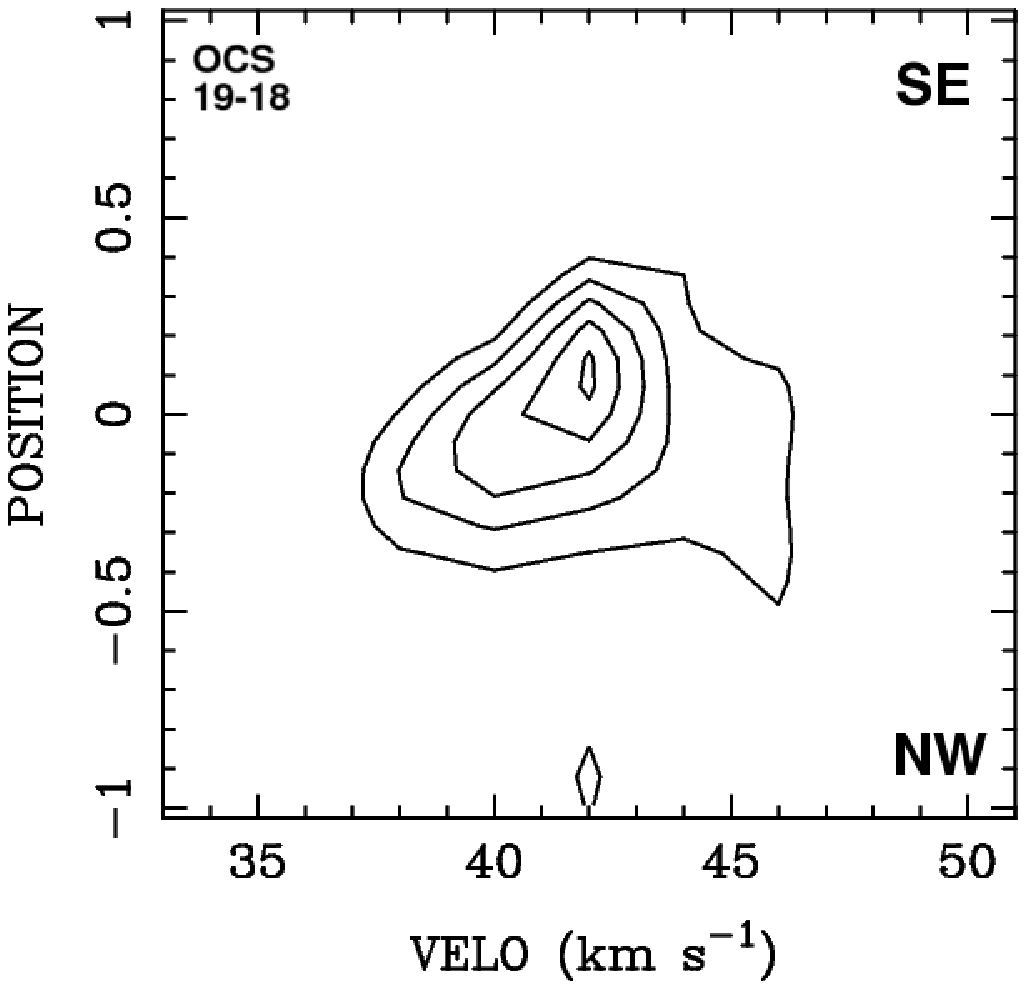} 

\plottwo{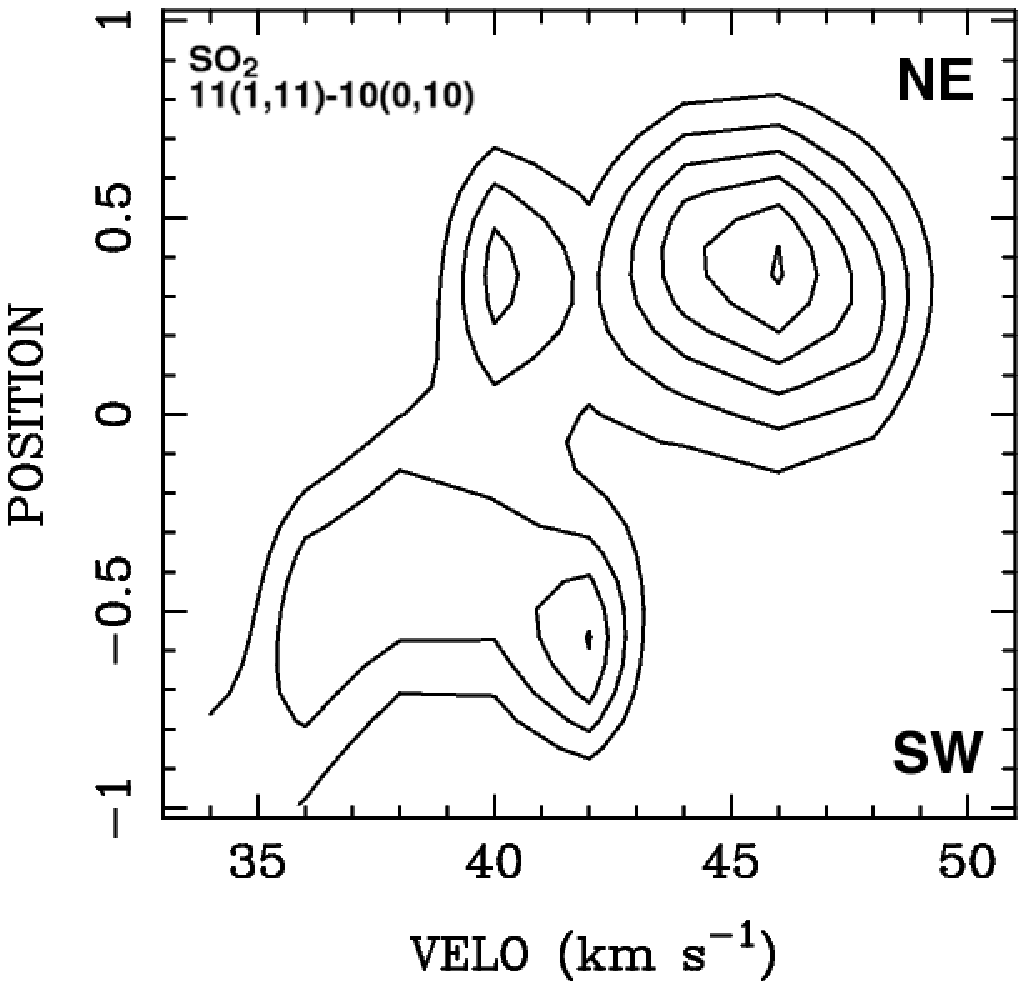}{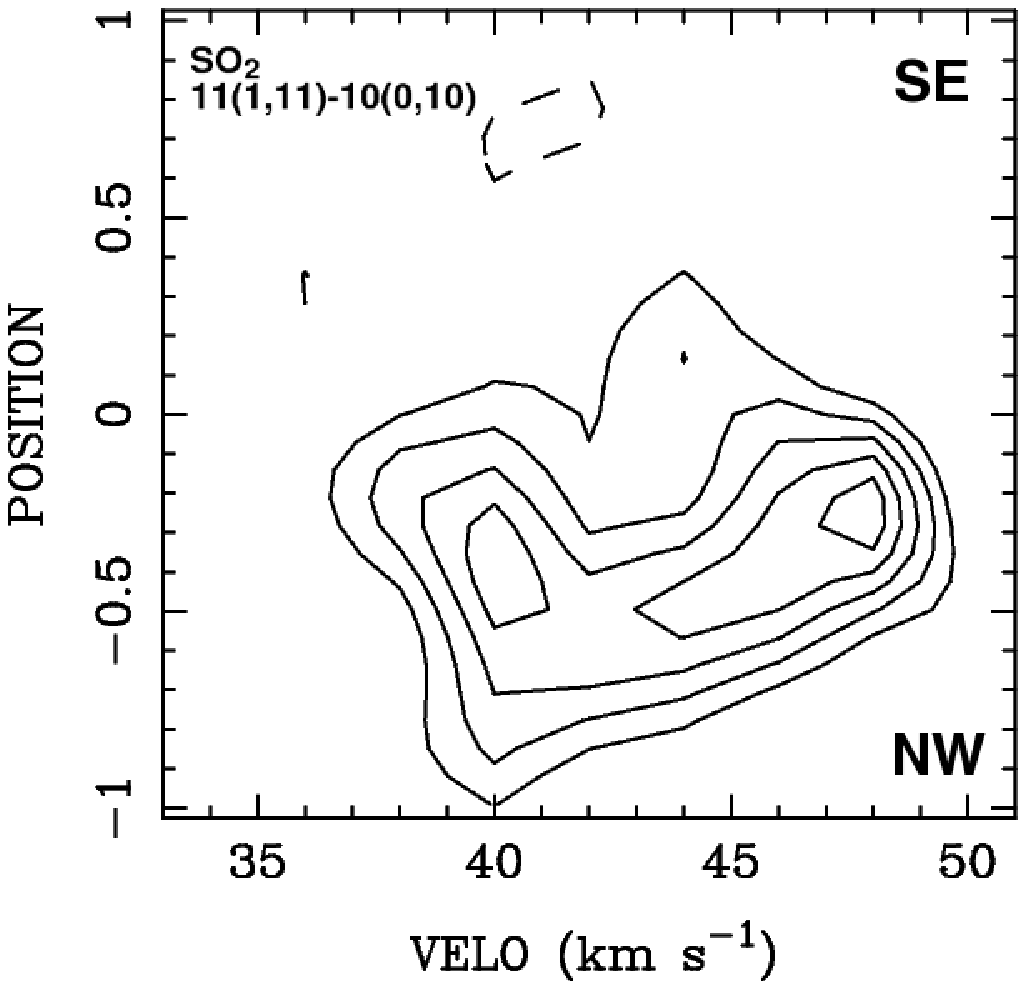} 

\plottwo{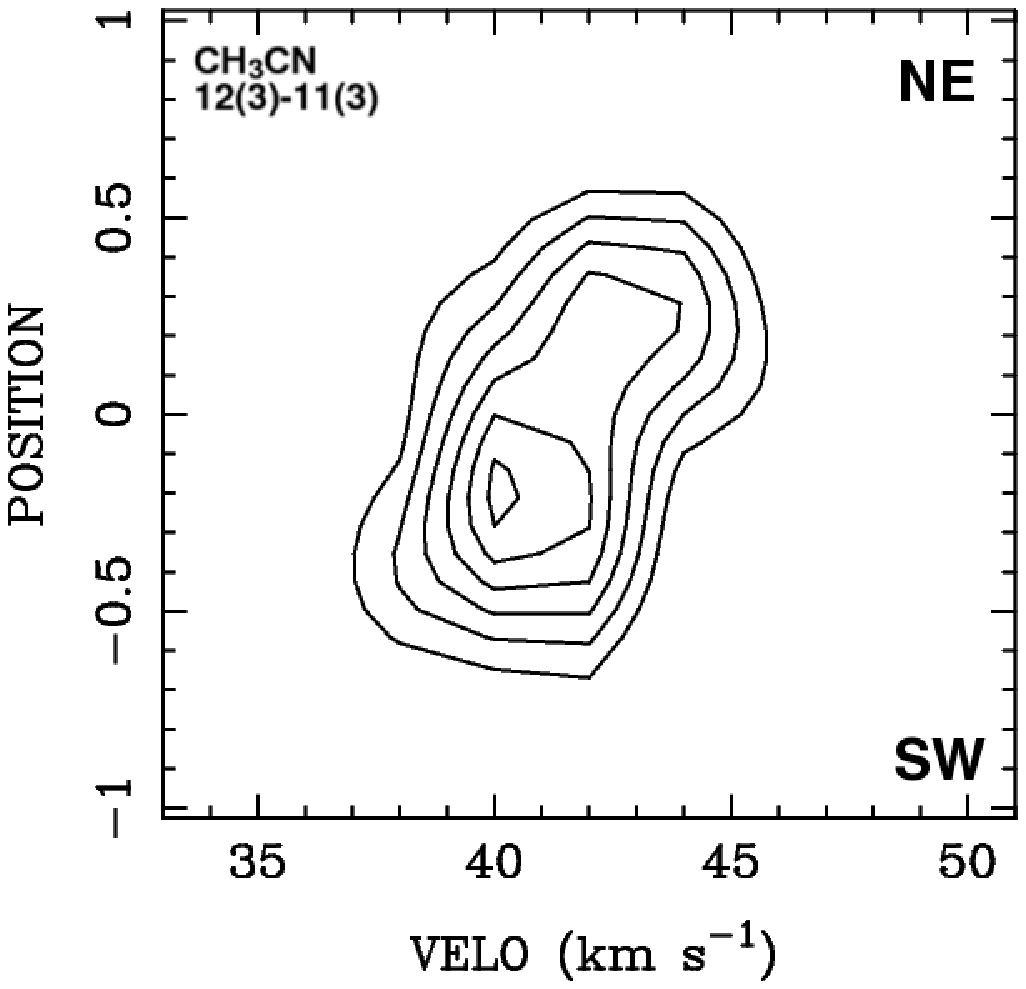}{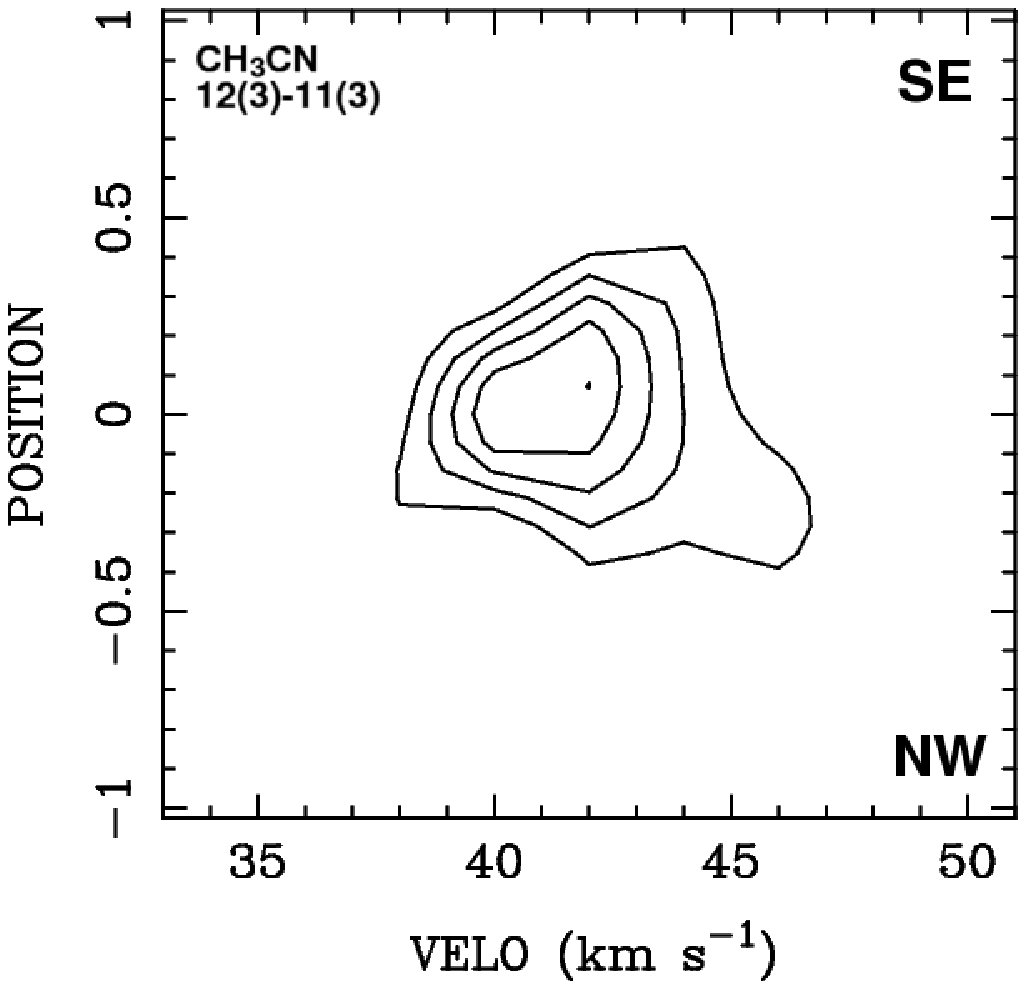}

\plottwo{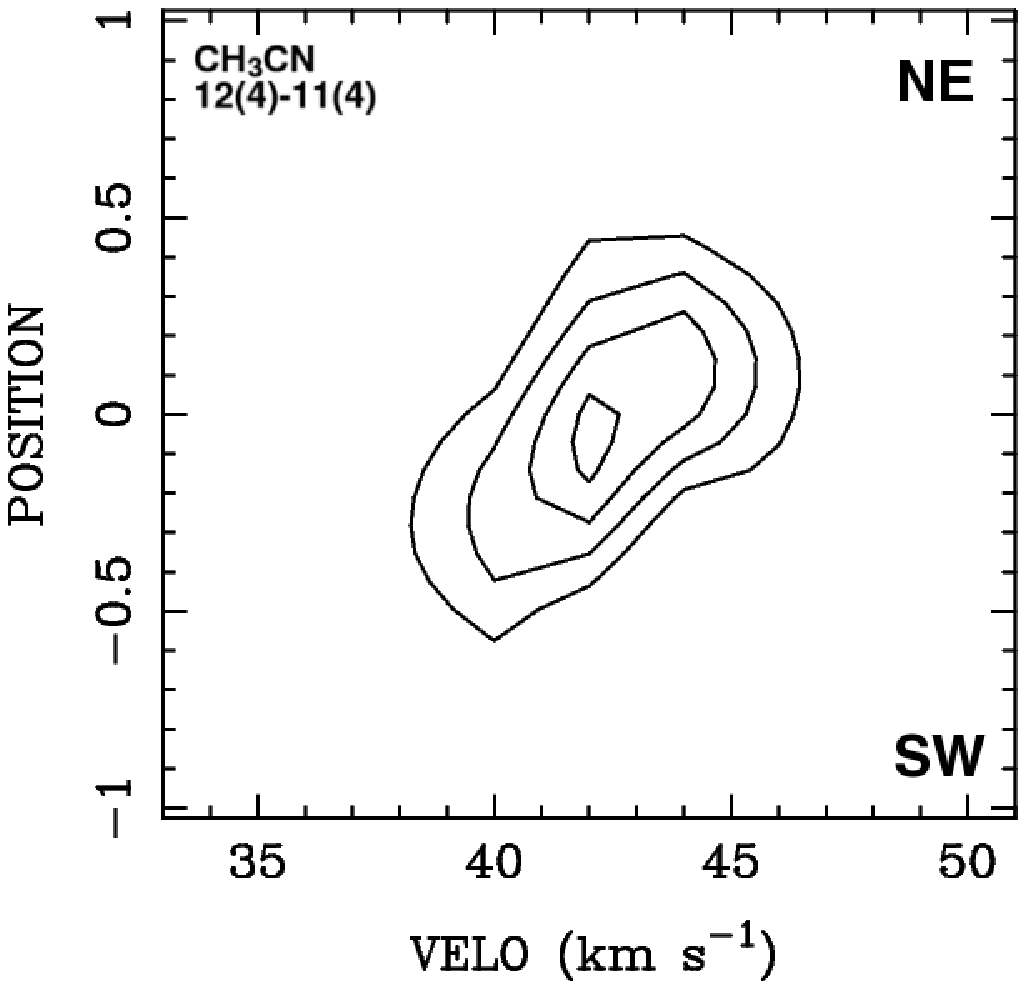}{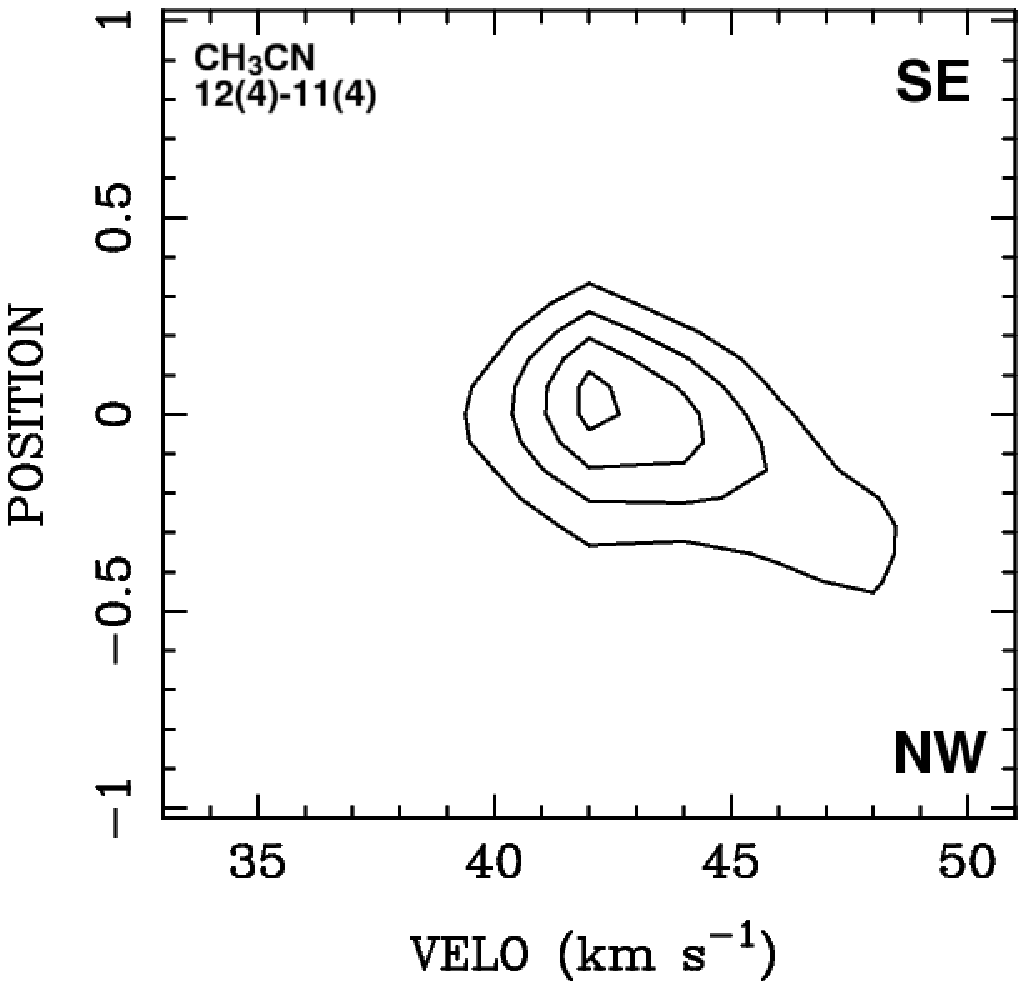}
\caption{
Position-velocity diagrams of the SMA-VEX observations. The {\it left} column is for a cut 
at PA$=45^\circ$ (SW-NE). The {\it right} column is at PA$=135^\circ$ (NW-SE). Cuts are centered 
at the position of the continuum peak of \HII region A. Contours are the same as in Fig. \ref{fig7a}. 
The velocity gradient interpreted as rotation from SW to NE is clearly seen in OCS and CH$_3$CN. 
The SO$_2$ features are more complex.}
\end{center}
\label{fig10}
\end{figure}

\clearpage

\begin{figure}
\figurenum{11}
\begin{center}
\includegraphics[angle=0,scale=0.5]{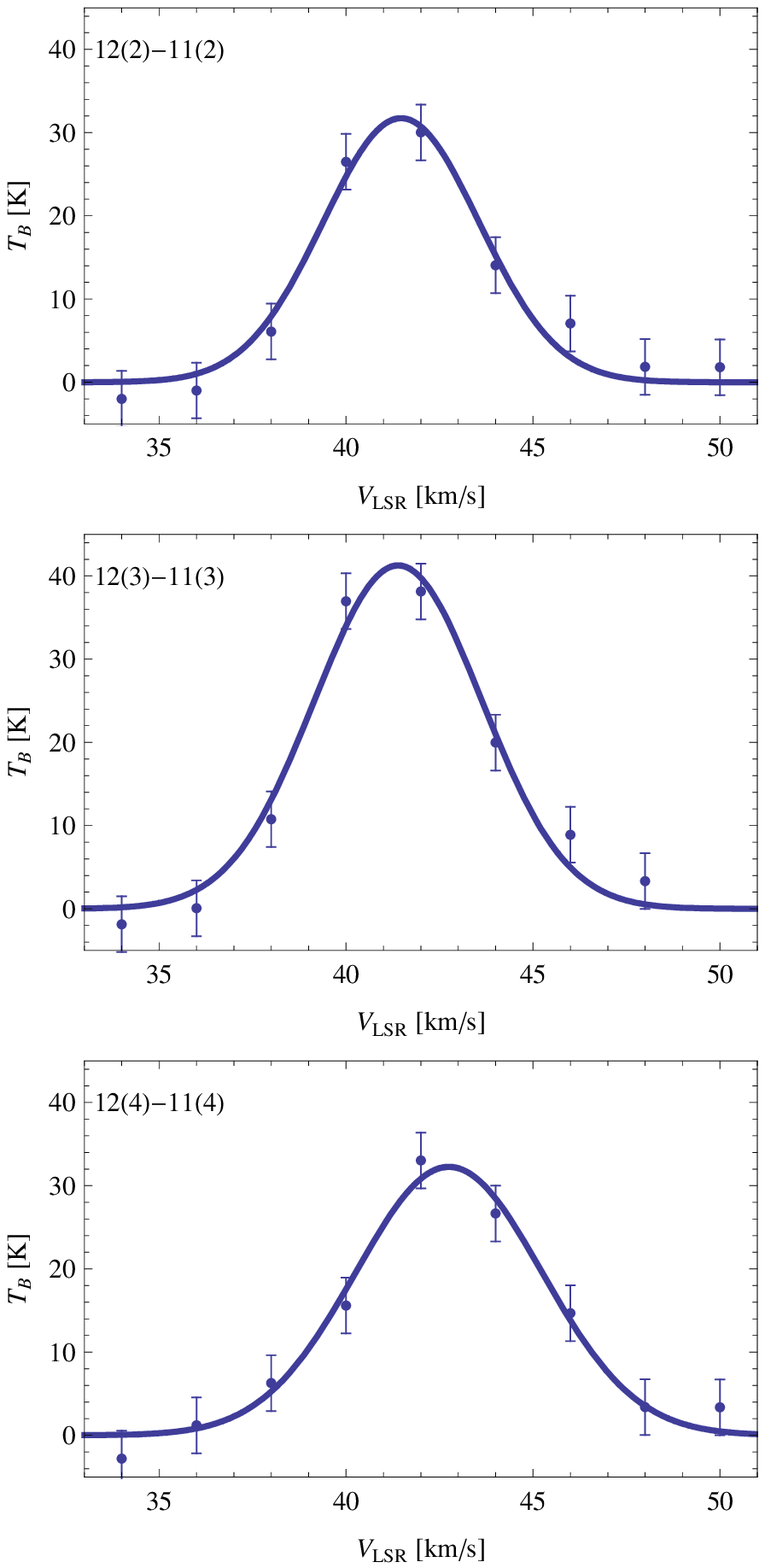}
\caption{
Spectra ({\it points}) and Gaussian fits ({\it lines}) to the CH$_3$CN $J=12-11$, $K=2,3,$ and 4 
lines at the position of \HII region A. 
Error bars denote the $1\sigma$ noise in the $2~\kms$-wide channels. 
}
\label{fig11}
\end{center}
\end{figure}

\begin{figure}
\figurenum{12}
\begin{center}
\includegraphics[angle=0,scale=0.53]{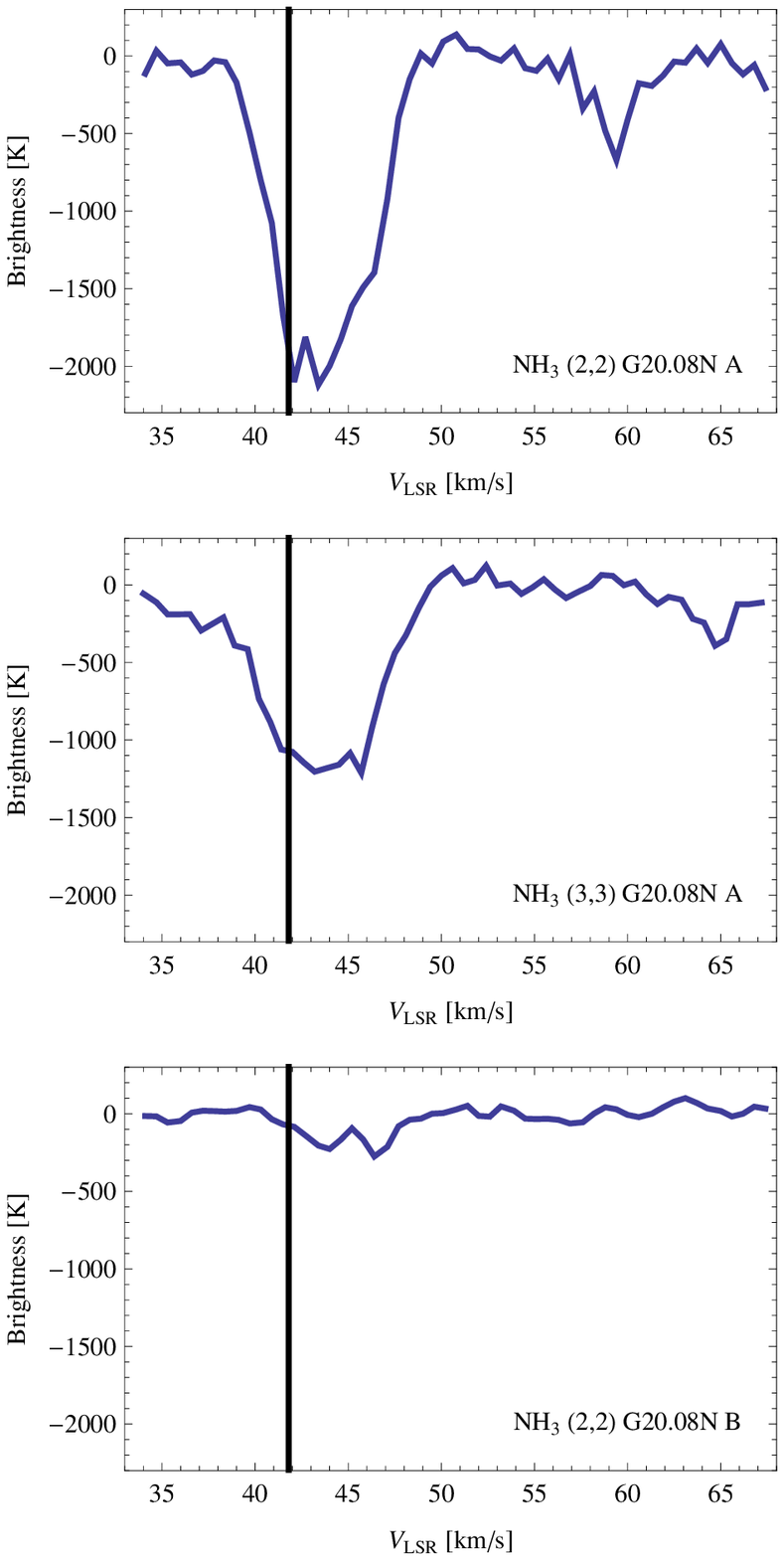}
\caption{
{\it Top} and {\it middle} panels: 
NH$_3$ (2,2) and (3,3) spectra toward the absorption peak of \HII region A 
in the VLA-BnA observations. 
The vertical  line marks the systemic velocity ($V_{sys} = 41.8~\kms$) 
of the molecular gas at scales comparable to the \HII region. The centers of the absorption lines 
are redshifted with respect to $V_{sys}$, indicating inflow of molecular gas toward \HII region A 
at small scales. The (3,3) absorption is broader (FWHM $\approx 6.5~\kms$) than the (2,2) 
(FWHM $\approx 5.6~\kms$), probably caused by larger motions closer to the center. 
{\it Bottom} panel: NH$_3$ (2,2) spectrum toward \HII region B. The absorption, considerably 
fainter than for \HII region A, is also redshifted. 
}
\label{fig12}
\end{center}
\end{figure}

\clearpage

\begin{figure}
\figurenum{13}
\epsscale{0.58}
\begin{center}
\plotone{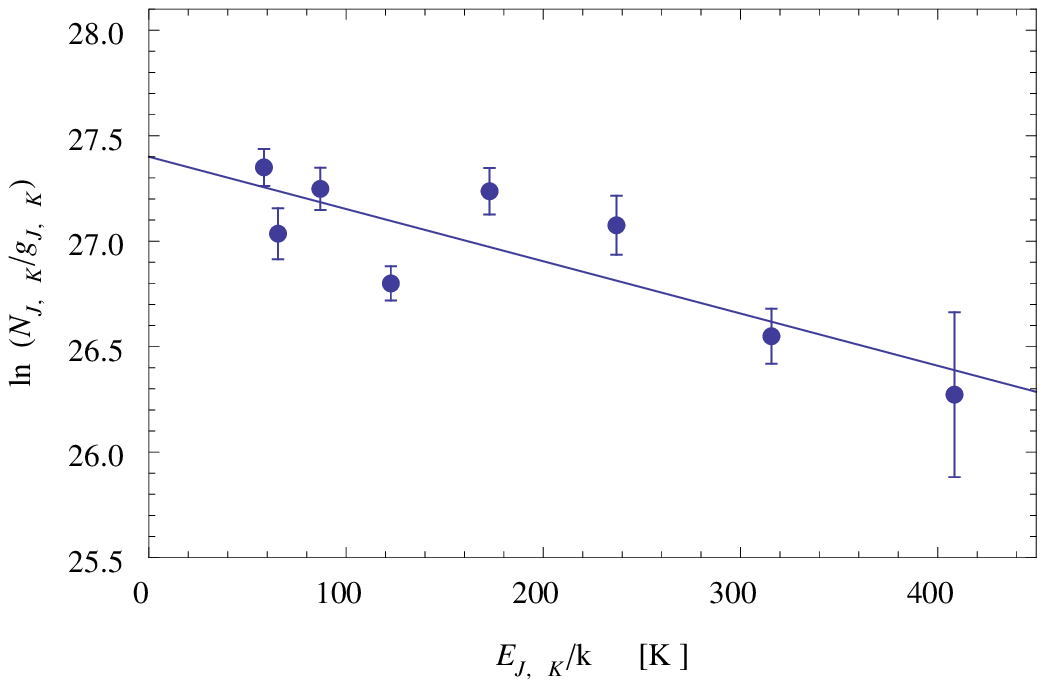}
\vspace{0.5cm}
\plotone{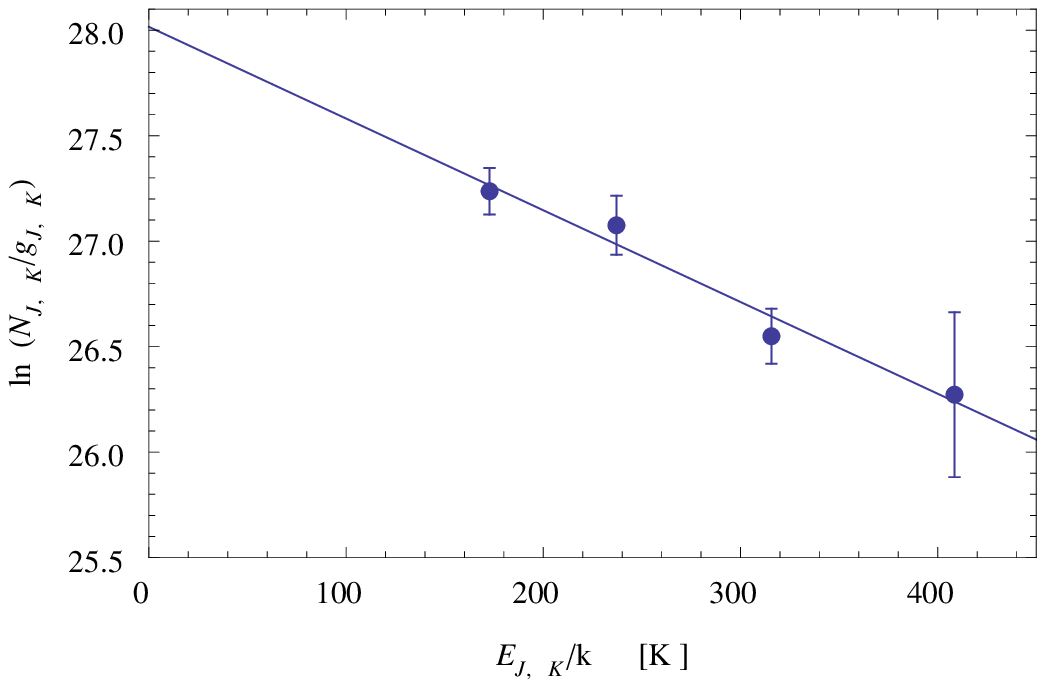}
\caption{
Rotation diagram for the CH$_3$CN $J=12-11$. 
The {\it top} panel is a fit of all the $K=0,...,7$ lines. The {\it bottom} panel uses only 
the $K=4,...,7$ lines, which have lower optical depths. 
Error bars are $3\sigma$ values. Lines are the linear fits to the data.}
\label{fig13}
\end{center}
\end{figure}

\begin{figure}
\figurenum{14}
\begin{center}
\includegraphics[angle=0,scale=0.48]{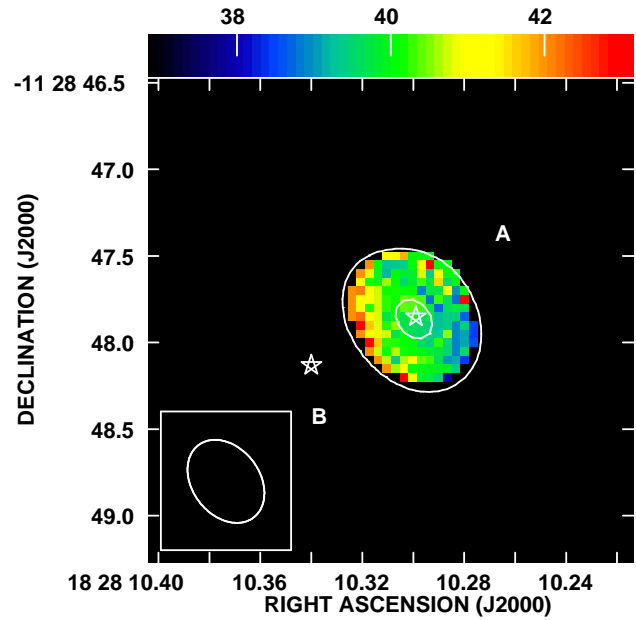}
\caption{
Velocity-integrated emission (moment 0, {\it contours}) and intensity-weighted mean velocity  
(moment 1, {\it color scale}) map of the H$30\alpha$ RRL emission toward G20.08N. 
Contours are at $7,50$ $\times$ $0.5~\jybkms$. Although the emission is unresolved 
at half power, the moment 1 map hints at the presence of a velocity gradient in the ionized gas 
similar to that seen in CH$_3$CN and OCS. 
}
\label{fig14}
\end{center}
\end{figure}

\clearpage

\begin{figure}
\figurenum{15}
\begin{center}
\includegraphics[angle=0,scale=0.9]{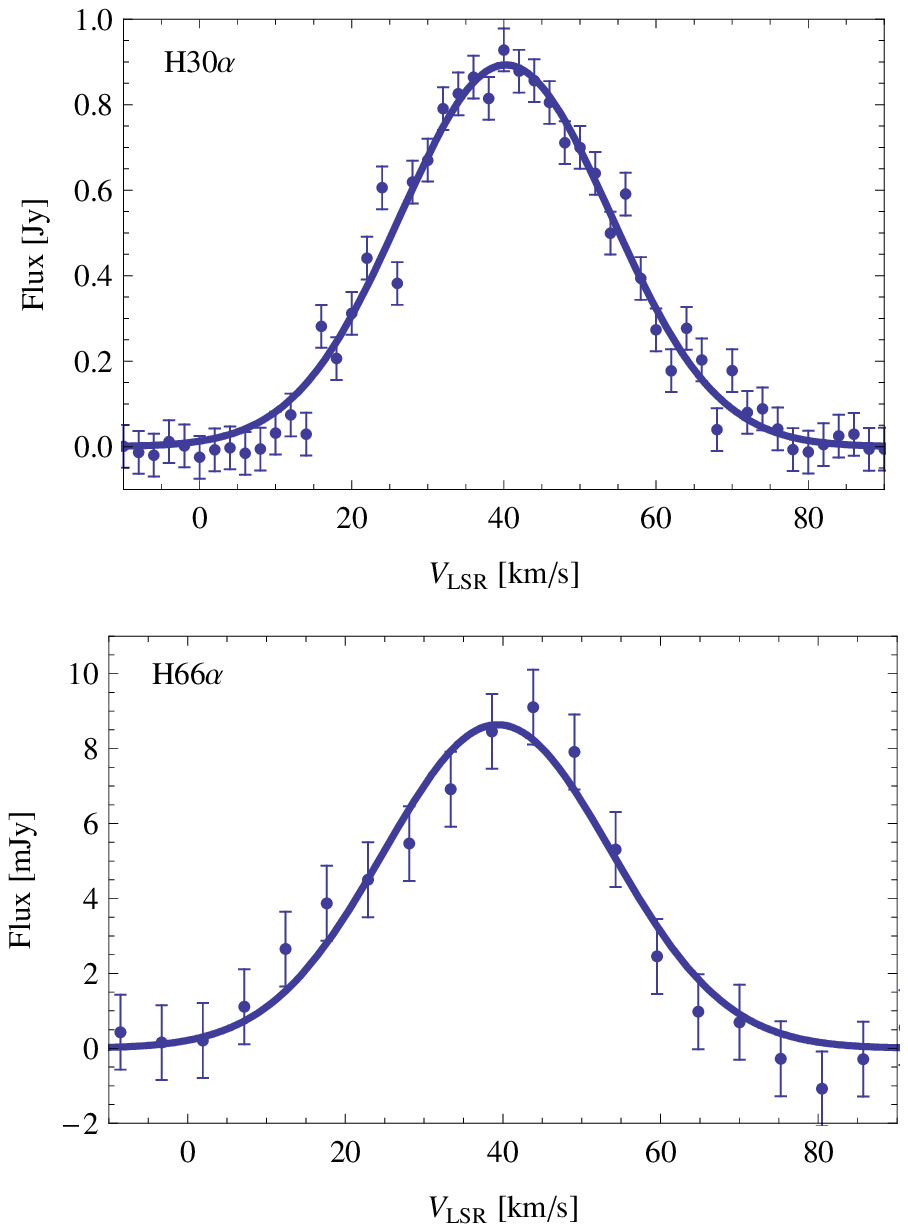}
\caption{
Spectra ({\it points}) and Gaussian fits ({\it lines}) to the H$30\alpha$ ({\it top} panel) and H$66\alpha$ 
({\it bottom} panel) lines toward G20.08N A. 
Error bars denote the $1\sigma$ noise in the channels. 
The channel spacing is $2~\kms$ for H$30\alpha$ and $5.2~\kms$ for H$66\alpha$. 
The flux was integrated over a 0.5\arcsec ~ square box centered on \HII region A. 
}
\label{fig15}
\end{center}
\end{figure}


\begin{figure}
\figurenum{16}
\begin{center}
\includegraphics[angle=0,scale=0.8]{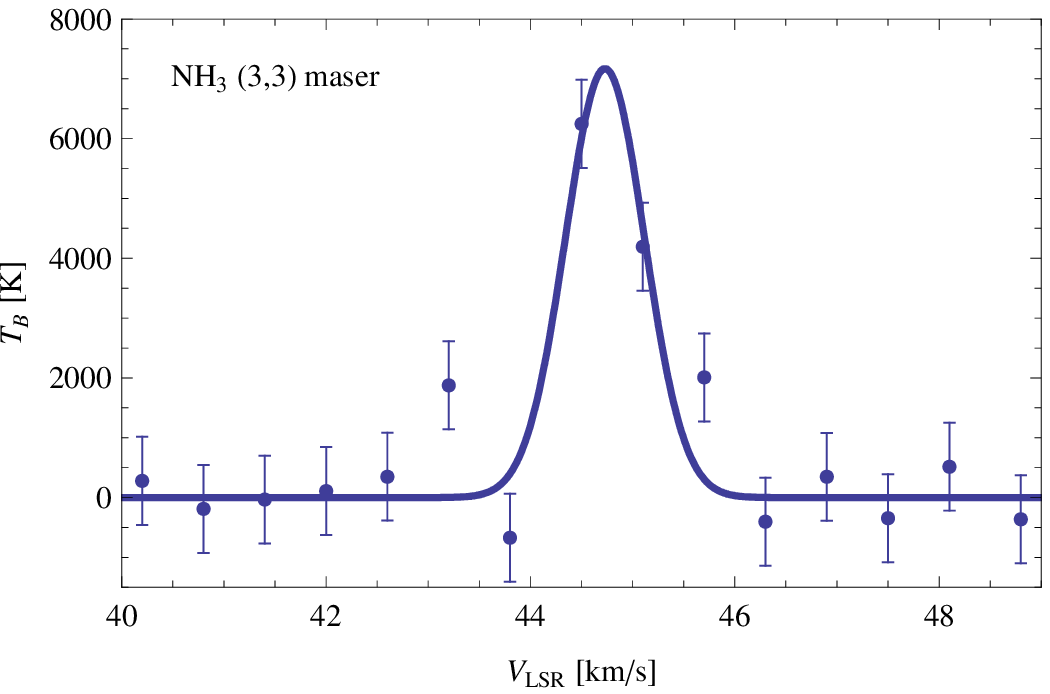}
\caption{
VLA-BnA spectrum of the NH$_3$ (3,3) maser spot toward G20.08N. The brightness temperature 
$T_B$ scale assumes an angular size equal to half the uniform-weighted beam 
dimensions.
}
\label{fig16}
\end{center}
\end{figure}

\end{document}